\documentclass[fleqn,usenatbib]{mnras}

\usepackage{newtxtext,newtxmath}

\usepackage[T1]{fontenc}

\DeclareRobustCommand{\VAN}[3]{#2}
\let\VANthebibliography\thebibliography
\def\thebibliography{\DeclareRobustCommand{\VAN}[3]{##3}\VANthebibliography}

\usepackage{graphicx}	
\usepackage{amsmath}	
\usepackage{hyperref}
\usepackage{multirow}
\usepackage{ulem}
\usepackage{comment}
\usepackage{xcolor}     

\newlength{\VSpaceBeforeTabBib}
\newlength{\VSpaceBeforeTabFoot}
\title[DCF Method Revisited]{The Davis-Chandrasekhar-Fermi Method Revisited}

\author[C.-Y. Chen et al.]{
Che-Yu Chen,$^{1,2}$\thanks{E-mail: cheyu.c@gmail.com}
Zhi-Yun Li,$^{2}$
Renato R. Mazzei,$^{2}$
Jinsoo Park,$^{3}$
Laura M. Fissel,$^{3}$
\newauthor
Michael C.-Y. Chen,$^{3}$
Richard I. Klein,$^{1,4}$
and Pak Shing Li$^{4}$
\\
$^{1}$Lawrence Livermore National Laboratory, Livermore, CA 94550, USA\\
$^{2}$Department of Astronomy, University of Virginia, Charlottesville, VA 22904, USA\\
$^{3}$Department of Physics, Engineering Physics and Astronomy, Queen’s University, Kingston, ON K7L 3N6, Canada \\
$^{4}$Department of Astronomy, University of California Berkeley, Berkeley, CA 94720, USA\\
}

\date{Accepted XXX. Received YYY; in original form ZZZ}

\pubyear{2022}

\begin{document}
\label{firstpage}
\pagerange{\pageref{firstpage}--\pageref{lastpage}}
\maketitle

\begin{abstract}
Despite the rich observational results on interstellar magnetic fields in star-forming regions, it is still unclear how dynamically significant the magnetic fields are at varying physical scales, because direct measurement of the field strength is observationally difficult.
The Davis-Chandrasekhar-Fermi (DCF) method has been the most commonly used method to estimate the magnetic field strength from polarization data. It is based on the assumption that gas turbulent motion is the driving source of field distortion via linear Alfv{\'e}n waves.
In this work, using MHD simulations of star-forming clouds, we test the validity of
the assumption underlying the DCF method by examining its accuracy in the real 3D space.  
Our results suggest that the DCF relation between turbulent kinetic energy and magnetic energy fluctuation should be treated as a statistical result instead of a local property.
We then develop and investigate several modifications to the original DCF method using synthetic observations, and propose new recipes to improve the accuracy of DCF-derived magnetic field strength. We further note that the biggest uncertainty in the DCF analysis may come from the linewidth measurement instead of the polarization observation, especially since the line-of-sight gas velocity can be used to estimate the gas volume density, another critical parameter in the DCF method.
\end{abstract}




\section{Introduction}

Magnetic fields have long been recognized to play a critical role in shaping the formation and evolution of molecular clouds and protostellar systems \citep{MO2007}, but definitive progress has been slow because the complete three-dimensional structure and strength of the magnetic field within molecular clouds cannot be directly probed observationally.
On one hand, the magnetic field strength along the line of sight can be derived via Zeeman splitting of molecular lines, but the measurements are notoriously difficult. Firm Zeeman detections thus remain sparse (see e.g.,~\citealt{Falgarone_Zeeman_2008,Troland_Zeeman_2008,Crutcher_2010}, or \citealt{Crutcher_review_2012} for a review).
On the other hand, polarized dust emission is generally thought to be a reliable tracer of the projected magnetic field direction on the plane of sky, 
because non-spherical grains tend to be oriented with their long axes perpendicular to the magnetic field lines (\citealt{Davis_Greenstein_1951}; also see review by \citealt{Lazarian_2007}). 
However, though dust polarization patterns have been successfully mapped at multiple scales from diffuse clouds to protostellar disks \citep[e.g.,][]{PlanckXIX, Hull_2013,Stephens_Nature_2014, Fissel_BLASTPol_2016,BISTRO_2017},
the 3D structure of magnetic field 
remains unknown.

With the advent of several new instruments in the last several years including {\it Planck} \citep[e.g.,][]{PlanckXIX}, BLASTPol \citep{Fissel_BLASTPol_2016}, JCMT/POL-2 \citep{BISTRO_2017}, and SOFIA/HAWC+ \citep{HAWC+}, the observational situation has improved drastically.
In particular, it is now possible to generate large number of polarization vectors in multi-scale observations with high-sensitivity polarimeters, which enables statistical examination on cloud polarization features. The statistical approach has proven to provide promising methods in theoretical studies \citep[e.g.,][]{Padoan_2001,FG08,Soler_HRO_2013}, and various models have been proposed to statistically approximate the observationally inaccessible properties of the magnetic field \citep[e.g.,][]{Chen_HRO_2016,Chen2019,King_2018,King19,Sullivan_2021}.

Despite the recent advances, 
the Davis-Chandrasekhar-Fermi (DCF) method \citep{Davis_1951,CF_1953} remains one of the most commonly used methods to estimate the magnetic field strength. The fundamental picture of the DCF method is to consider the propagation of Alfv{\'e}n waves along the originally-uniform magnetic field. In this scenario, any distortion of the field lines correlates directly to the gas motions (see e.g.,~Fig.~\ref{fig:defAlf}). Therefore, by measuring the field distortion (usually traced by the dispersion of polarization angles) and the turbulent strength (traced by linewidth), the field strength can be estimated by assuming 1) energy balance holds between the gas kinetic and magnetic energy fluctuations (i.e.,~$\delta E_K \approx \delta E_B$), and 2) the gas turbulent motion is isotropic in 3D.
While this method has been tested with numerical simulations \citep[e.g.,][]{Heitsch01,OSG2001,FG08} and applied to observational data at various scales \citep[e.g.,][]{Girart_IRAS4A_2006,Pillai_IRDC_2015,PlanckXXXV,Pattle_BISTRO_2017,Kwon_DCFdisk_2019},
it should be noted that the highly restrictive assumptions of the DCF method about the gas motions and magnetic field geometry severely compromise its accuracy, especially for star-forming regions that are self-gravitating \citep[see e.g.,][]{OSG2001}. 

Several theoretical studies have been conducted towards characterizing the uncertainties of the DCF method, with most of the efforts being focused on investigating the cancellation effect in observed polarization angle dispersion through either the beam convolution \citep[e.g.,][]{Zweibel1990,MG91} or the integration along the line of sight \citep[e.g.,][]{FG08,Hildebrand_09,Houde09,Cho_Yoo_2016,Cho_2019,ST_DCF_2021}.
However, as we shall argue in this work, the most severe uncertainty in the DCF method when applying to real observations is likely the hydrodynamic properties of the gas, not the polarization measurement. 
In fact, as pointed out in the recent work by \cite{PSLi_2021}, the interstellar MHD waves are likely nonlinear, and there exist many other modes that do not satisfy the Alfv{\'e}n relation, which is a prerequisite of the DCF method.
{Several recent theoretical works have also investigated the correlation between turbulent kinetic energy and the perturbed component of magnetic energy using fully 3D simulation data, and suggested that the applicability of the DCF-assumed scenario is limited \citep[see e.g.,][]{Skalidis_ST_2021, Beattie_2022}.}
If the DCF equation does not hold in 3D, its application on projected 2D data is questionable, unless proper calibration can be provided to increase the accuracy.

In this paper, we revisit the fundamental picture behind the DCF method using fully-3D MHD simulations (Sec.~\ref{sec::DCF3D}), and investigate the applicability of the simplified DCF scenario toward star-forming gas in 2D synthetic observations (Sec.~\ref{sec:synobs}).
We first examine the balance between turbulent kinetic energy and the perturbed component of the magnetic energy in the 3D space using MHD simulations, which was assumed to be in equipartition in the DCF scenario. We then use the linear Alfv{\'e}n wave relation (the foundation of the DCF method) to derive the ``unperturbed'' field $\mathbf{B}_0$ for the simulated cloud in 3D (Sec.~\ref{sec:Alftest}), which is a pure mathematical result assuming the DCF equation holds everywhere. 
We also discuss why the mean field could be considered as the background, ``unperturbed'' field when conducting the DCF analysis (Sec.~\ref{sec:B0approx}-\ref{sec:statcorr}).
In Sec.~\ref{sec:synobs} we discuss possible corrections and modifications for the DCF method when applying on 2D, projected observational data, which we test and compare in Sec.~\ref{sec::test}. In Sec.~\ref{sec:2Ddisp} we propose a new method to estimate the correction factor for the DCF equation from the cancellation effect on polarization angles along the line of sight.
We summarize our conclusions in Sec.~\ref{sec:sum}.

\section{The 3D DCF Relation}
\label{sec::DCF3D}

\begin{table}
 \begin{center}
  \caption{Summary of the simulation models considered in this study. These simulations were originally reported in \protect\cite{CO15} and \protect\cite{Chen2019}. Values here represent the averaged values in the shock-compressed regions in these colliding-cloud simulations (see Sec.~\ref{sec::DCF3D} for more detailed description on the model setups).
  {Here, $\langle\delta\phi\rangle$ represents the angle dispersion of magnetic field direction in 3D space.}}
  \label{tab:models}
  \begin{tabular}{lcccc}
    \hline
    \hline
    model & $B_{\rm 3D}$ & plasma & $v_{\rm rms}$ &
    $\langle \delta\phi \rangle$ \\
    & ($\mu$G) & $\beta$ & (km\,s$^{-1}$) & $(^\circ)$\\
    \hline
    L1 & 74 & 0.15 & 0.76 & 12 \\
    L5 & 13 & 0.14 & 0.99 & 20 \\
    L10 & 12 & 0.05 & 1.87 & 19 \\
    L20 & 20 & 0.03 & 2.15 & 25 \\
    \hline
  \end{tabular}
  \end{center}
\end{table}

{
In this section, we describe our investigation toward the DCF correlation, and demonstrate our analysis using 3D MHD simulations. 
We used the set of simulations reported in \cite{Chen2019}, which are cloud-scale, core-forming 3D MHD simulations.\footnote{Note that model L1 represents a slightly different scenario than other models. This particular simulation was designed to follow a local shock-compressed region (1\,pc in size) within a molecular cloud, and thus the initial gas density is higher (1000\,cm$^{-3}$) and has relatively strong magnetic field ($\sim 75\,\mu$G) comparing to typical cloud-scale properties ($\sim 10-100$\,cm$^{-3}$ and $\sim 10\,\mu$G). Since the focus of this study is to test the DCF method under various circumstances, we include model L1 with other models in our analysis, though we note that L1 is a relatively extreme case for cloud-scale gas and magnetic field structures.} 
The core formation activities in these clouds are induced by turbulent convergent flows, which compresses the originally diffuse gas ($\sim 10-100$\,cm$^{-3}$) to create dense, post-shock regions within which dense structures and cores form \citep[for more detailed information, see][]{CO14,CO15,Chen_HRO_2016}. For our analysis in this study, we only consider the post-shock regions in the simulations because they resemble better the physical properties of the observed star-forming molecular clouds. A summary of the basic cloud properties of these models is in Table~\ref{tab:models}. Generally speaking, models L1 and L20 represent the cases with relatively the strongest magnetic field and turbulence, respectively, and models L5 and L10 represent the moderately magnetized cases, with model L10 being more turbulent than L5 (see e.g.,~the values of plasma $\beta$ and the rms velocity in Table~\ref{tab:models}). 
}

\subsection{The Original DCF Method}

\begin{figure}
\begin{center}
\includegraphics[width=\columnwidth]{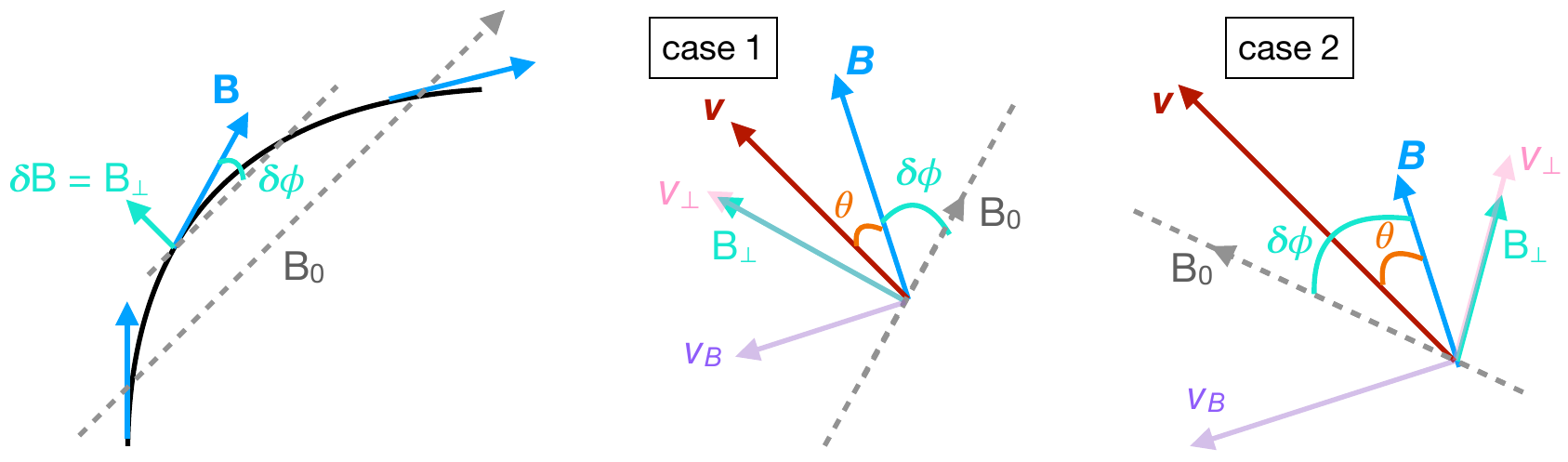}
\caption{\small {\it Left:} Sketch of the perturbed magnetic field $\delta B$ in the direction perpendicular to the unperturbed magnetic field direction induced by an Alfv{\'e}n wave running along the ``initial'' field $B_0$. {\it Middle and right panels:} Illustrating the geometric relation between the gas velocity $v$ and magnetic field $B$, as well as their perturbation components $v_\perp$ and $\delta B = B_\perp$, following the linear Alfv{\'e}n wave equation (see Eq.~(\ref{eq:vpvA}) and Sec.~\ref{sec:Alftest}). }
\label{fig:defAlf}
\end{center}
\end{figure}

Based on the assumption that the fluctuation of gas kinetic energy is equal to the fluctuation of magnetic energy, $\delta E_K = \delta E_B$,
the original DCF method envisioned the interplay between gas turbulence and magnetic field as a transverse Alfv{\'e}n wave propagating through a background magnetic field $B_0$.
The gas velocity thus induces small deviation of the magnetic field, $\delta B$, from the initial, unperturbed field $B_0$. 
Since 
{we restrict our analysis to Alfv{\'e}nic fluctuations,} 
only the perpendicular component of the gas velocity with respect to the initial field $B_0$ is effective here, and the perturbed component of the magnetic field is perpendicular to $B_0$ ({$B_\parallel = 0$,} $B_\perp = \delta B$; see the left panel of Fig.~\ref{fig:defAlf}). 
This suggests the relation between the gas velocity and the perturbed magnetic field should follow
\begin{equation}
    \frac{\delta E_K}{\delta E_B} = 1 = \frac{\rho v_\perp^2/2}{B_\perp^2/(8\pi)} = \frac{4\pi\rho v_\perp^2}{B_\perp^2}.\label{eq:dEratio}
\end{equation}
The perturbed magnetic field $B_\perp$ and gas velocity $v_\perp$ thus satisfy
\begin{equation}
v_\perp =V_A \frac{\delta B}{B_0},\label{eq:vpvA}
\end{equation}
where $V_A \equiv \sqrt{B_0^2/4\pi\rho}$ is the Alfv{\'e}n velocity in the pre-perturbation medium. 
This implies, if there is a well-defined, small amplitude Alfv{\'e}n wave running along the ``initial'' field, $B_0$, it is possible to derive the magnitude of $B_0$ using the DCF equation:
\begin{equation}
    B_0 = \sqrt{4\pi\rho}\frac{v_\perp}{\delta B/B_0} \approx \sqrt{4\pi\rho}\frac{v_\perp}{\delta\phi},
    \label{eq:DCF3D}
\end{equation}
where $\rho$, $v_\perp$, and $\delta\phi$ can be estimated from observations (see Sec.~\ref{sec:synobs}).
Note that there is no prerequisite (e.g.,~must be constant) on $B_0$ to satisfy Eq.~(\ref{eq:vpvA}), and thus we shall not simply treat it as the average field over the interested region before further justifications (see Sec.~\ref{sec:B0approx}).

\subsection{The Initial Magnetic Field in the DCF Method}
\label{sec:Alftest}

If the assumption of small-amplitude Alfv{\'e}n wave is valid, Eqs.~(\ref{eq:dEratio})-(\ref{eq:DCF3D}) should hold everywhere in the medium .
This allows an inference of the unperturbed background field direction $\hat{\mathbf{B}_0}$ in each location when $\mathbf{B}$ and $\mathbf{v}$ vectors are known. As illustrated in Fig.~\ref{fig:defAlf}, in the plane made by $\mathbf{B}$ and $\mathbf{v}$ vectors, let $\theta$ be the angle between $\mathbf{B}$ and $\mathbf{v}$ vectors and $\delta\phi$ the angle between $\mathbf{B}$ and $\mathbf{B}_0$ {(all angles are positive)}, we have either (note that $\mathbf{B}_0$ cannot be in between $\mathbf{B}$ and $\mathbf{v}$)
\begin{equation}
\sqrt{4\pi \rho} v\sin(\delta\phi+\theta) = B\sin(\delta\phi)
\end{equation}
for case 1 ($\mathbf{B}_0$ on $\mathbf{B}$ side), or
\begin{equation}
\sqrt{4\pi \rho} v\sin(\delta\phi-\theta) = B\sin(\delta\phi)
\end{equation}
for case 2 ($\mathbf{B}_0$ on $\mathbf{v}$ side). 
Note that here we require $\delta\phi < 90^\circ$, which is a requirement for the DCF method to be applicable.
This is because, for the DCF method to work accurately, there must exist such $\mathbf{B}_0$ as the original, pre-perturbation field. Since the perturbed field $\mathbf{B}$ is the combination of $\mathbf{B}_0$ and the perturbed component $\mathbf{\delta B}$ from gas turbulence, and since only the velocity component perpendicular to $\mathbf{B}_0$ could bend the field line, we have $\mathbf{B} = \mathbf{B}_0 + \mathbf{\delta B}$ with $\mathbf{B}_0 \perp \mathbf{\delta B}$, and the angle between $\mathbf{B}$ and $\mathbf{B}_0$ must be less than $90^\circ$.

We would like to point out that, case 1 represents the ``traditional'' view of the DCF relation that when the deviation of $\mathbf{B}$ from $\mathbf{B}_0$ is not large, or in general, when the gas turbulence is sub-Alfv{\'e}nic. On the other hand, case 2 is also physically correct with $B_0 \ll B_\perp \lesssim B$, which represents the situation when the gas turbulence is much more energetic comparing to the initial magnetic energy (i.e.,~super-Alfv{\'e}nic).\footnote{{Because in case 2, $\sqrt{4\pi\rho} v = \sqrt{4\pi\rho} v_\perp/\sin(\theta+\delta\phi) > B_\perp/\sin(\delta\phi) = B$. Thus, locations that satisfy the scenario of case 2 are locally super-Alfv{\'e}nic. Similarly, case 1 represents sub-Alfv{\'e}nic locations.}}
Another criterion for case 2 to be valid is $\delta\phi > \theta$, which is a numerical requirement but also provide the natural limit that $\theta$ must be smaller than $90^\circ$ in this scenario.


We can now solve $\delta\phi$ following the derivations below:
\begin{align}
    &\sqrt{4\pi\rho}\,v \sin (\delta\phi\pm\theta) = B \sin\delta\phi, \\
    &x \equiv \sin\delta\phi,\ \ A\equiv \frac{\sqrt{4\pi\rho}\, v}{B},\\
    & A = \frac{x}{x\cos\theta \pm \sin\theta\sqrt{1-x^2}},\\
    &\sin^2\delta\phi = x^2 = \frac{A^2\sin^2\theta}{1-2A\cos\theta+A^2}.
\end{align}
This gives $\delta\phi$ (and thus the direction of $\mathbf{B}_0$) at every location.
Note that the derived formula of $\delta\phi$ is the same for both cases.
Since $\mathbf{B}_0$, $\mathbf{B}$, and $\mathbf{v}$ must be on the same plane,
we can therefore solve for $v_\perp$ and $B_\perp$:
\begin{equation}
    v_\perp = v\sin(\delta\phi\pm\theta),\ \ \ B_\perp = B\sin\delta\phi.
\end{equation}
When applying on simulation data, we adopt case 1 as the default solution, and we only use the solution from case 2 when 
\begin{equation}
    \frac{1}{2}\rho (v\cos\theta)^2 > \frac{B^2}{8\pi}, \ \ \ {\rm and} \ \ \ \delta\phi > \theta.
\end{equation}

\begin{figure*}
\begin{center}
\includegraphics[width=\textwidth]{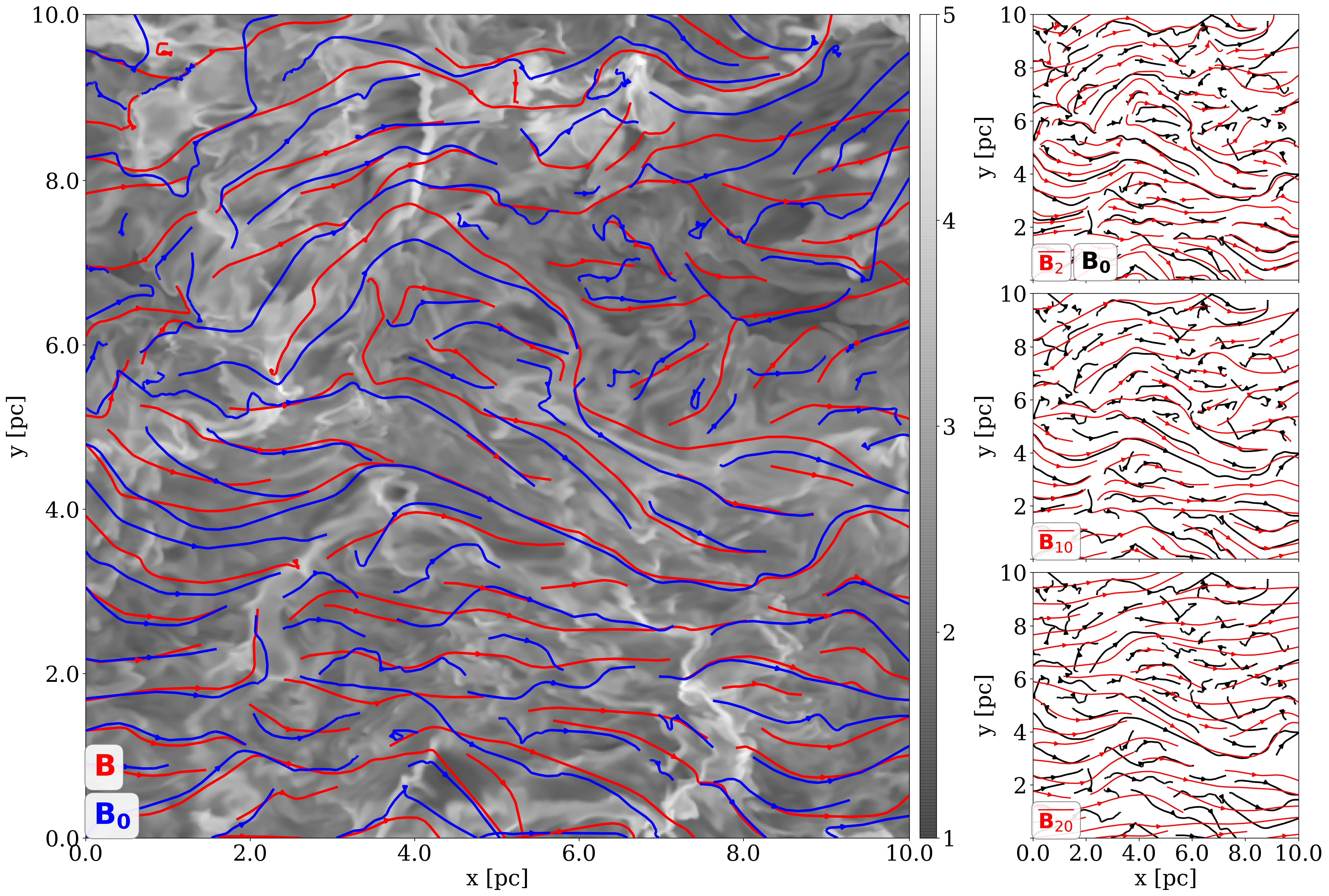}
\vspace{-.2in}
\caption{{\it Left:} An example of comparing the reconstructed $\mathbf{B_0}$ (the $x-y$ components only; {\it blue streamlines}) and the original magnetic field $\mathbf{B}$ (also the $x-y$ components only; {\it red streamlines}), overplotted on a slice of gas density (in $\log (n / [{\rm cm}^{-3}])$, grayscale) cut through the mid-plane in the post-shock layer of a turbulent colliding flow simulation (model L10 in \citealt{Chen2019}; see Table~\ref{tab:models}).
.{\it Right:} Comparisons between the ``unperturbed'' field $\mathbf{B_0}$ ({\it black streamlines}, same in all three panels) and the vector-averaged fields $\overline{\mathbf{B}_s}$ ({\it red streamlines}) for $s=2$ ({\it top}), $10$ ({\it middle}), and $20$ ({\it bottom}), for the same density slice shown in the left. The gas structure is not shown here for clarity.  }
\label{fig:Alftest}
\end{center}
\end{figure*}

By defining $\mathbf{v_B}$ as the vector in the same plane of $\mathbf{B}_0$, $\mathbf{B}$, and $\mathbf{v}$ and is perpendicular to $\mathbf{B}$ (see Fig.~\ref{fig:defAlf}), we have
\begin{equation}
    \mathbf{v_B} = \mathbf{v} - \left(\mathbf{v}\cdot\hat{\mathbf{B}}\right)\hat{\mathbf{B}},
\end{equation}
and we can derive the direction of $\mathbf{B_0}$ (denoted as the unit vector $\hat{\mathbf{B}}$):
\begin{equation}
    \hat{\mathbf{B}_0} = \hat{\mathbf{B}}\cos\delta\phi \mp \hat{\mathbf{v_B}}\sin\delta\phi.
\end{equation}
{Since $\mathbf{B} = \mathbf{B_0} + \mathbf{\delta B}$ and $\mathbf{B_0} \perp \mathbf{\delta B}$, we know the amplitude of $\mathbf{B_0}$ should be $B_0 = B\,\cos\delta\phi$. However,}
we note that only the direction of $\mathbf{B}_0$ is important here, because that is what we need to derive $B_\perp$ for the DCF analysis.
Fig.~\ref{fig:Alftest} illustrates an example of the derived $\hat{\mathbf{B}_0}$ from a cloud-scale, core-forming simulation \citep[model L10 in][]{Chen2019}. 
Also note that, as long as we can determine $\hat{\mathbf{B}_0}$ properly, the DCF relation is applicable everywhere even in super-Alfv{\'e}nic gas, in contrast to the commonly-considered assumption that the DCF relation only holds in sub-Alfv{\'e}nic regime{, which have in fact been challenged by recent numerical studies \citep[see e.g.,][]{JHLiu_DCF_2021,Skalidis_ST_2021,Beattie_2022}. }

\subsection{Approximation of $\hat{\mathbf{B}_0}$}
\label{sec:B0approx}

\begin{figure*}
\begin{center}
\includegraphics[width=0.48\textwidth]{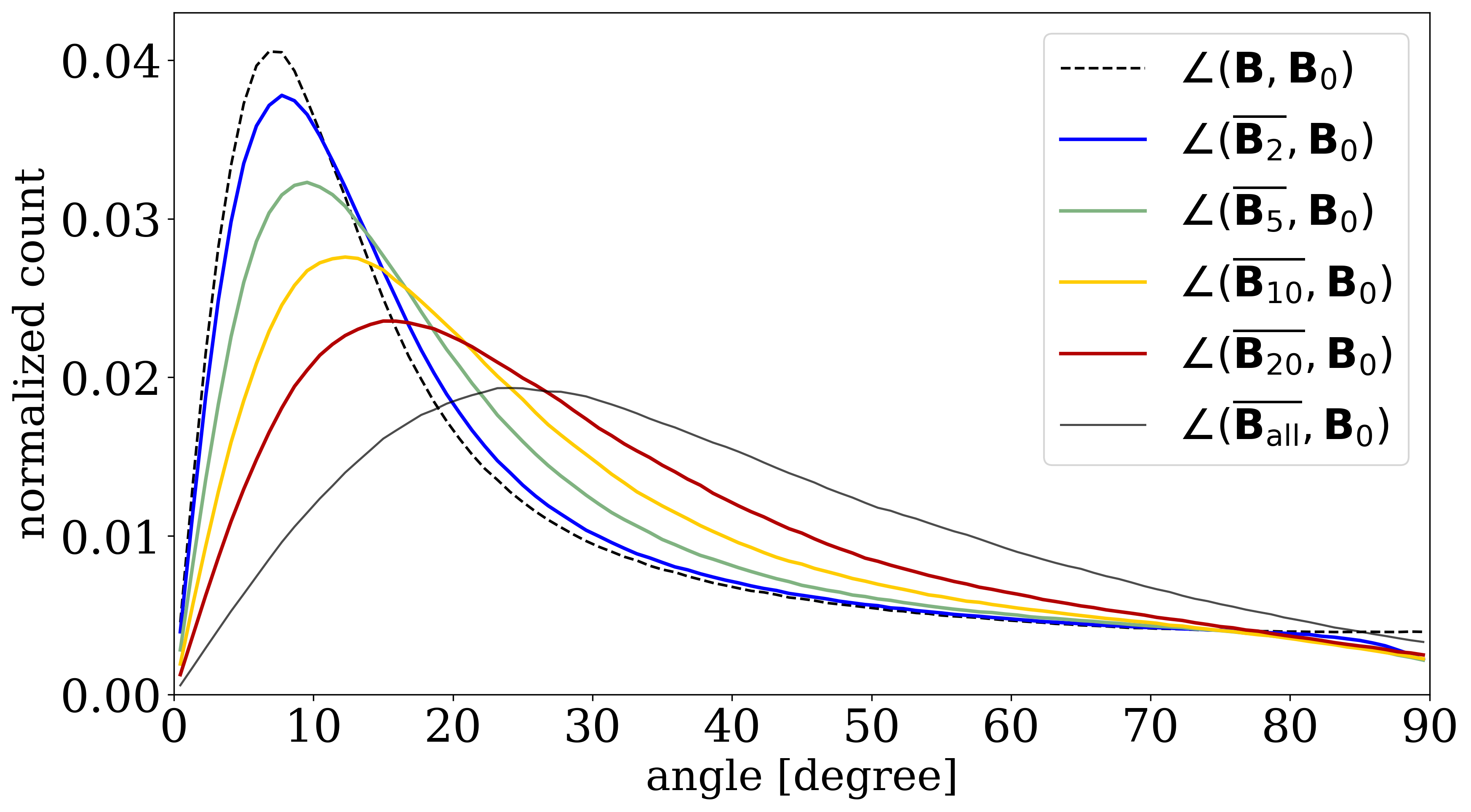}
\includegraphics[width=0.48\textwidth]{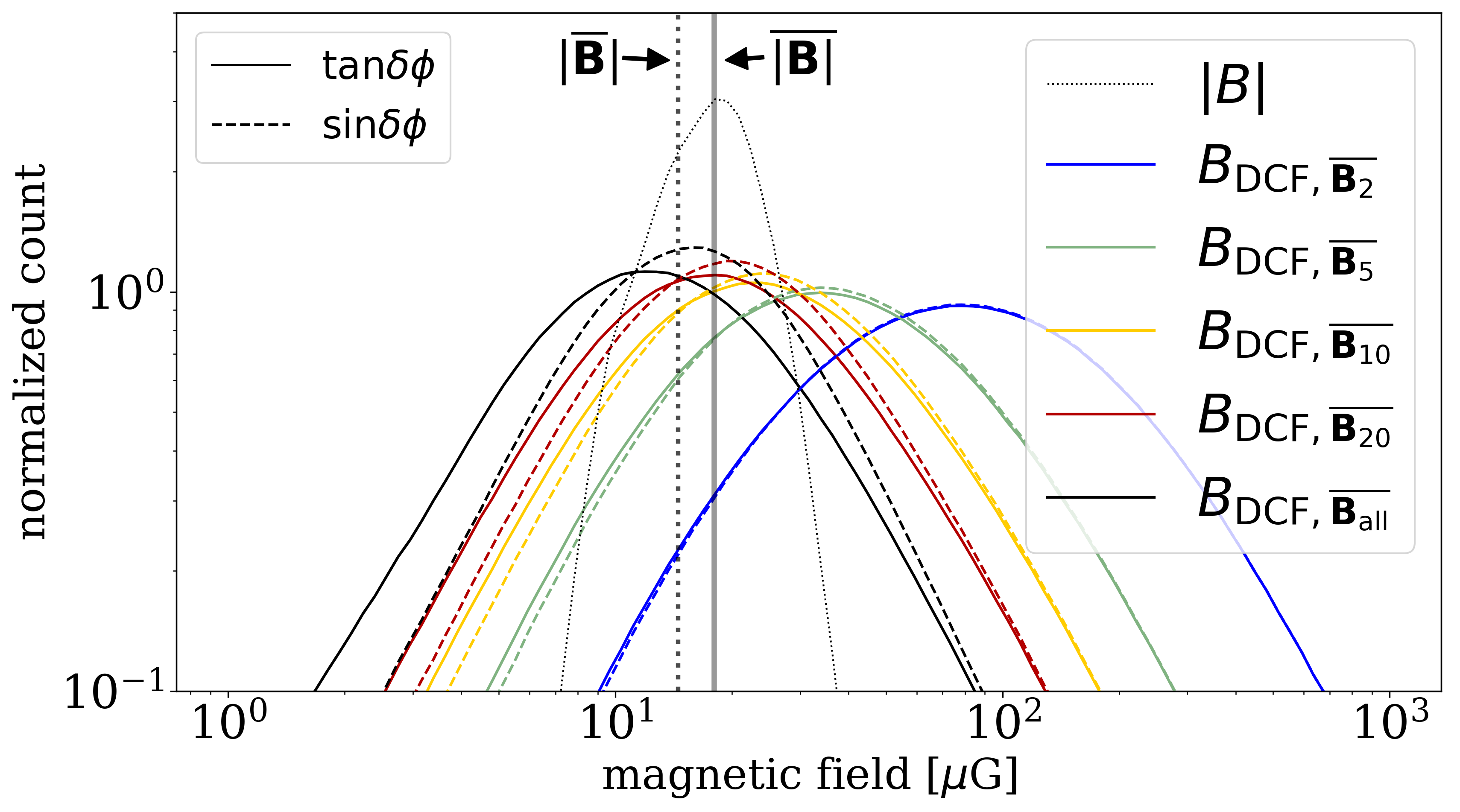}
\caption{\small Comparing $\overline{\mathbf{B}_s}$, $s = 2,\ 5,\ 10,\ 20$ ($\mathbf{B}$ averaged over $5^3$, $11^3$, $21^3$, and $41^3$ cells) by showing histograms of $\angle(\overline{\mathbf{B}_s}, \mathbf{B_0})$ ({\it left}) and DCF-derived magnetic field strength ({\it right}). Though $\overline{\mathbf{B}_s}$ seems to be a good approximate of the direction of $\mathbf{B_0}$ (i.e., small relative angle) statistically when $s$ is small (i.e., averaged locally), only the $B_{\rm DCF}$ derived from large-scale averaged $\overline{\mathbf{B}_s}$ gives good estimates of the averaged field strength. In addition, we found that using $\sin\delta\phi =  B_\perp/B$ ({\it dashed curves}) provides more accurate results than the commonly-adopted $\tan\delta\phi = B_\perp/|\overline{\mathbf{B}}|$ ({\it solid curves}). The gray vertical line on the right panel marks the mean field strength calculated from the scalar mean ({\it solid line}) and vector-average ({\it dotted line}).}
\label{fig:AlftestBavg}
\end{center}
\end{figure*}

\begin{figure*}
\begin{center}
\includegraphics[width=\textwidth]{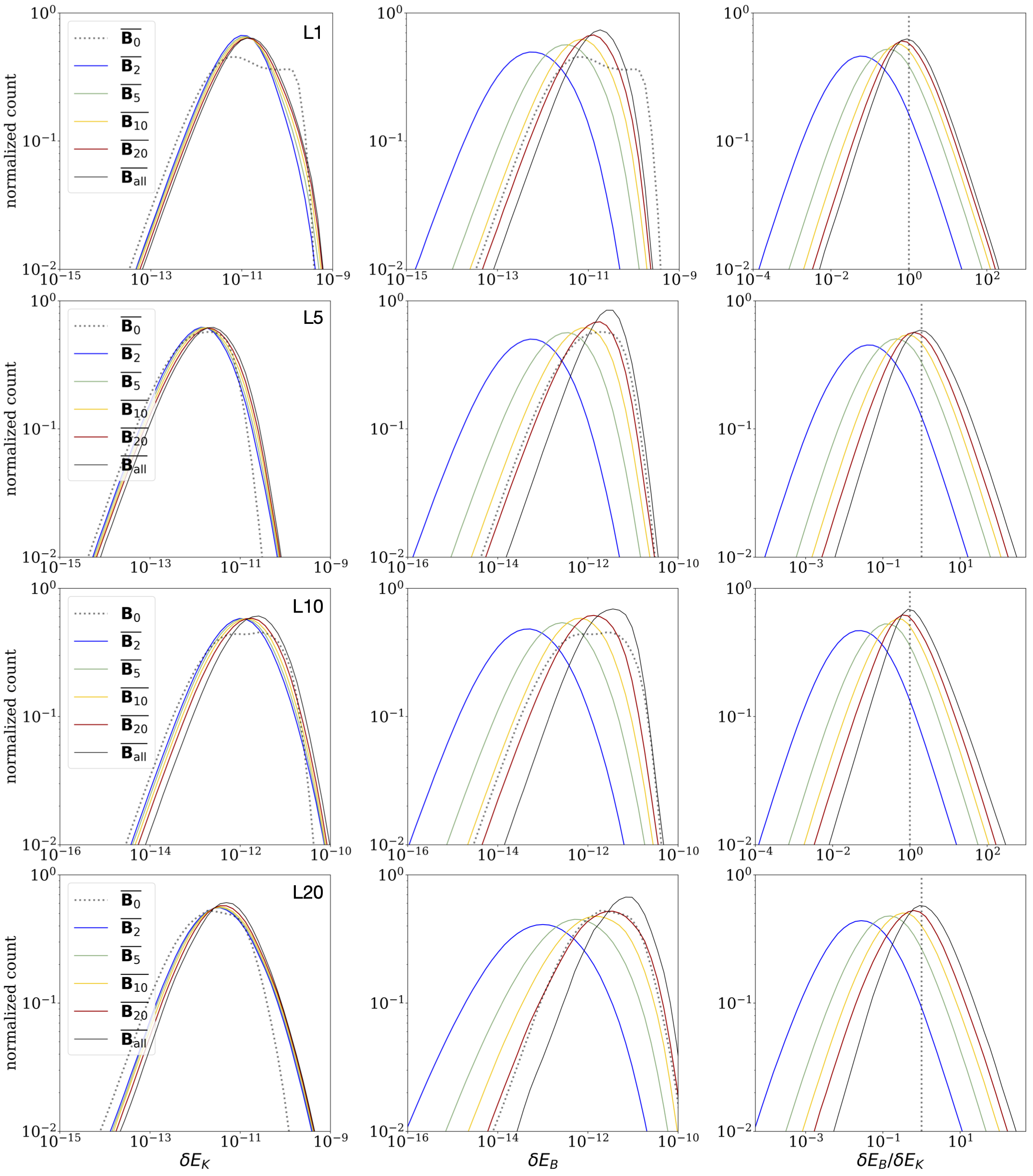}
\vspace{-.2in}
\caption{\small Comparing energy ratio $\delta E_B/\delta E_k \equiv (B_\perp^2/(8\pi))/(\rho v_\perp^2/2)$ with $B_\perp$, $v_\perp$ derived with $\overline{\mathbf{B}_s}$ as $\mathbf{B_0}$ {for all 4 models considered in this work (see Table~\ref{tab:models})}. When averaging over large scales, $\overline{\mathbf{B}_s}$ is able to make $\delta E_B/\delta E_k$ peak near unity, so the most probable value of the corresponding DCF-derived field strength thus roughly agrees with the mean field strength in the simulations (also see Fig.~\ref{fig:AlftestBavg}). }
\label{fig:AlftestdEratio}
\end{center}
\end{figure*}

As demonstrated in the previous section, the accuracy of the DCF method depends on the accuracy of the measurement of the direction $\hat{\mathbf{B}_0}$.
However, the full derivation of $\hat{\mathbf{B}_0}$ requires knowing the angle between $\mathbf{v}$ and $\mathbf{B}$, which cannot be probed in observations. 
Since theoretically $\mathbf{B_0}$ represents the ``unperturbed'' field,
a straightforward alternative is to consider the vector-average of $\mathbf{B} = B_x \hat{\mathbf{x}} + B_y \hat{\mathbf{y}} + B_z \hat{\mathbf{z}}$ over a given scale as an approximation of $\mathbf{B_0}$:
\begin{equation}
     \overline{\mathbf{B}_s}  = \overline{{B_{x,}}_s} \hat{\mathbf{x}} + \overline{{B_{y,}}_s} \hat{\mathbf{y}} + \overline{{B_{z,}}_s} \hat{\mathbf{z}},
\label{eq:B_n}
\end{equation}
where $\overline{{B_{x,}}_s}$, $\overline{{B_{y,}}_s}$, $\overline{{B_{z,}}_s}$ represent averaged $B_x$, $B_y$, and $B_z$ over the chosen scale, denoted as $s$.
Using the same simulation shown in Fig.~\ref{fig:Alftest} (model L10 in \cite{Chen2019}), we consider the averaging scale to be $\pm s$ simulation cells ($dx = 0.02$\,pc for this model), 
i.e.,~we calculate $\mathbf{B}_s$ following Eq.~(\ref{eq:B_n}) by averaging over a $(2s+1)^3$ volume centered at each cell.

To see how $\overline{\mathbf{B}_s}$ depends on scales and whether we can use $\overline{\mathbf{B}_s}$ to approximate $\hat{\mathbf{B}_0}$,
we calculate the angle difference between $\overline{\mathbf{B}_s}$ and $\hat{\mathbf{B}_0}$ utilizing the full 3D information of the simulation data. 
The cases of $s = 2,\ 5,\ 10,\ 20$ are shown in Fig.~\ref{fig:AlftestBavg} (left panel). Also included are the comparisons between the local field $\mathbf{B}$ (i.e.,~$s=0$) and the ``unperturbed'' field $\mathbf{B_0}$ as well as the total vector-averaged field $\overline{\mathbf{B_{\rm all}}}$ among the entire simulation domain ($512^3$ cells or $10^3$\,pc$^3$; see Fig.~\ref{fig:Alftest}).
Our results show that, statistically, the angle difference is smaller when the averaged field $\overline{\mathbf{B}_s}$ is derived within a smaller volume,
which implies that the direction of $\overline{\mathbf{B}_s}$ could only be a good approximation of $\hat{\mathbf{B}_0}$ locally (i.e.,~for small $s$).
%

We further use the derived $\overline{\mathbf{B}_s}$ to calculate $B_\perp$ and $v_\perp$, and use those values to calculate the DCF-derived magnetic field strength, $B_{{\rm DCF}, \overline{\mathbf{B}_s}}$, following Eq.~(\ref{eq:DCF3D}):
\begin{equation}
    B_{{\rm DCF}, \overline{\mathbf{B}_s}} = \sqrt{4\pi\rho}\frac{v_\perp}{B_\perp/|\overline{\mathbf{B}_s}|}.
\end{equation}
The results are plotted in Fig.~\ref{fig:AlftestBavg} (right panel).
Interestingly, when looking at the probability distribution of the DCF-derived field strength using the direction of $\overline{\mathbf{B}_s}$ as $\hat{\mathbf{B}_0}$, it is the large-scale average $\overline{\mathbf{B}_{20}}$, or the whole-box average $\overline{\mathbf{B}_{\rm all}}$, that has its most probably value around the actual mean field value.\footnote{Note that there are two ways to define the mean field strength: the absolute value of the vector-averaged field $|\overline{\mathbf{B}}|$, or the scalar-average of the magnitude of the field $\overline{|\mathbf{B}|}$. Both are indicated in the right panel of Fig.~\ref{fig:AlftestBavg} (vertical lines in the plot).} 
In contrast, the DCF method tends to overestimate the field strength when the reference field is averaged locally (small $s$ for $\overline{\mathbf{B}_s}$).
This is not surprising, because the locally-averaged magnetic field is more tightly correlated with the local field direction (see Fig.~\ref{fig:AlftestBavg}, left panel), which means the dispersion angle $\delta\phi = \angle(\mathbf{B}, \overline{\mathbf{B}_s})$ tends to be small, and thus the derived $B_{\rm DCF, 3D}$ tends to be large.

The right panel of Fig.~\ref{fig:AlftestBavg} suggests that the vector-averaged field direction over a large scale could be adopted as the ``unperturbed'' field direction $\hat{\mathbf{B_0}}$ in the DCF equation to provide a good estimate of the mean field strength within the selected region, even though the direction of such averaged field may not agree with $\hat{\mathbf{B_0}}$ everywhere (see the right column of Fig.~\ref{fig:Alftest}). 
We note that this can be explained by the distribution of $\delta E_K \equiv \rho v_\perp^2/2$ and $\delta E_B \equiv B_\perp^2/(8\pi)$ with respect to the averaged magnetic field direction, which is analogous to the original DCF assumption Eq.~(\ref{eq:dEratio}).
{Note that the turbulent kinetic energy $\delta E_K$ considered here is the total kinetic energy from the perpendicular component of the gas velocity with respect to various reference magnetic field direction ($\overline{\mathbf{B}_s}$ and $\mathbf{B_0}$) within the cloud.}
As shown in Fig.~\ref{fig:AlftestdEratio}, while the distribution of $\delta E_K$ does not vary much when the referencing direction changes, the distribution of $\delta E_B$ shifted towards larger values when the referencing field direction is derived over larger scales. This is expected, because as we mentioned above, locally-averaged field direction tends to give smaller $B_\perp$, and thus $\delta E_B$ is smaller. 
When we use the entire simulation domain to calculate the reference field direction $\overline{\mathbf{B}_{\rm all}}$, the corresponding $\delta E_B/\delta E_K$ peaks around 1, consistent with the prerequisite of the DCF method (Eq.~(\ref{eq:dEratio})), thus provide the most accurate estimate of the field strength in Fig.~\ref{fig:AlftestBavg}).

\begin{figure*}
\begin{center}
\includegraphics[width=0.85\textwidth]{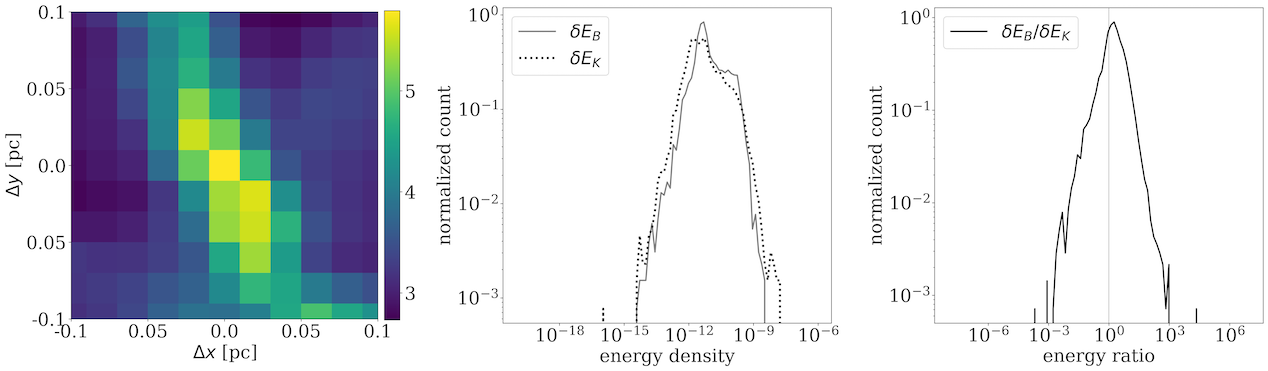}
\includegraphics[width=0.85\textwidth]{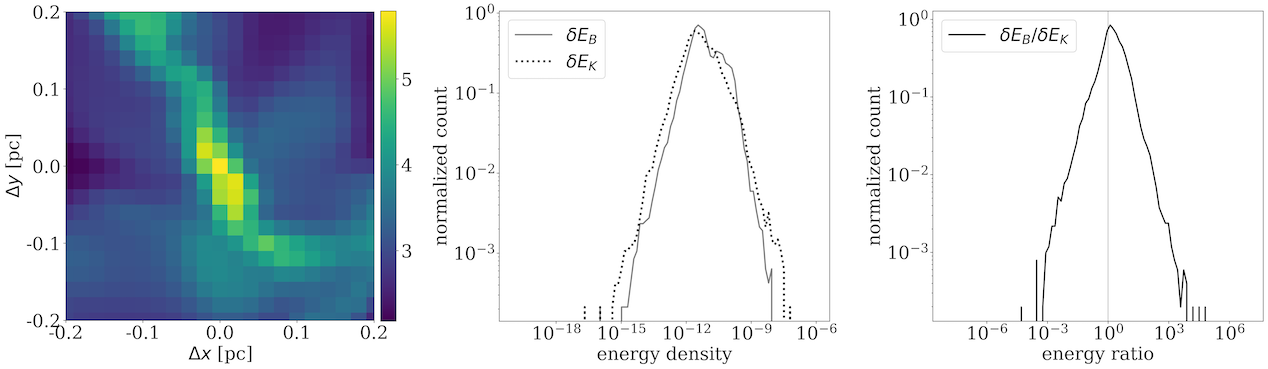}
\includegraphics[width=0.85\textwidth]{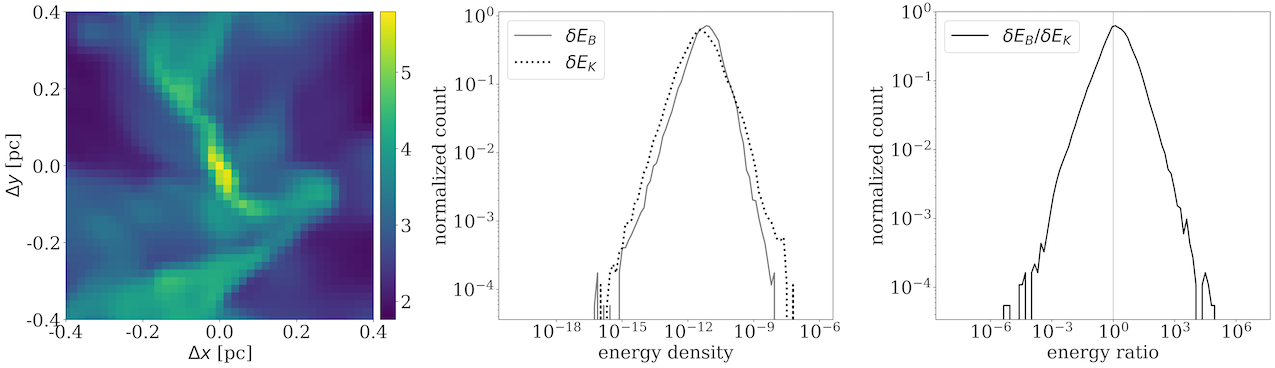}
\includegraphics[width=0.85\textwidth]{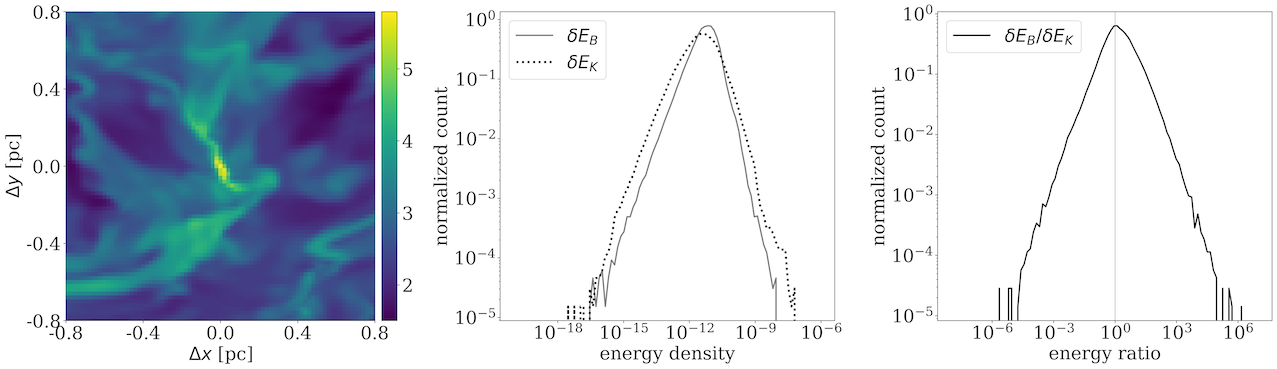}
\includegraphics[width=0.85\textwidth]{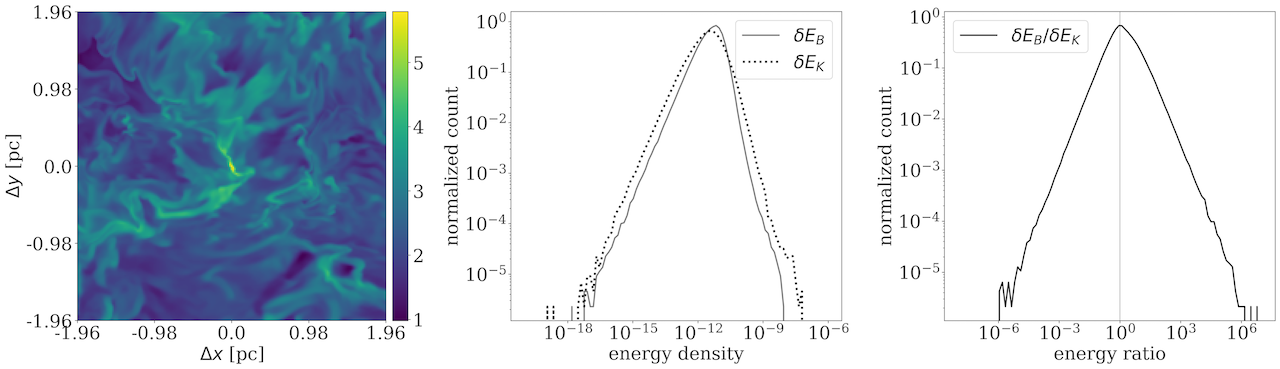}
\caption{\small Equipartition between $\delta E_K$ and $\delta E_B$ in dense core and surrounding regions covering different scales (small to large from top to bottom). {\it Left column:} Slice plot of the gas density $\log (n / [{\rm cm}^{-3}])$ demonstrating the corresponding gas structures. {\it Middle and right columns:} The probability distributions of $\delta E_B$, $\delta E_K$ ({\it middle}), and $\delta E_B/\delta E_K$ ({\it right}) using the vector-averaged field over the entire domain shown in the left column as the reference field. When including more diffuse gas in the background, the distribution of $\delta E_B/\delta E_K$ becomes more log-normal with a peak near 1 (equipartition). }
\label{fig:Alftest_coremaps}
\end{center}
\end{figure*}

We further note that this property of equipartition between the most probable values of $\delta E_K$ and $\delta E_B$ holds in sub-regions as well, and even around dense cores. 
Fig.~\ref{fig:Alftest_coremaps} illustrates a series of the same analysis (using $\overline{\mathbf{B}_{\rm all}}$ as the reference field direction) applied on regions of different scales centered at the same dense core. While we clearly see the shape of the $\delta E_B/\delta E_K$ distribution becomes more log-normal when including more background clouds around the dense core (from top to bottom rows of Fig.~\ref{fig:Alftest_coremaps}), the $\delta E_B/\delta E_K$ distribution from the smallest scale that we tested (top row) still peaks very close to 1, which suggests the DCF analysis we performed in Fig.~\ref{fig:AlftestBavg} may still be a good approximation even in the immediate surrounding of a dense core.


We would like to point out that the equipartition in $\delta E_B$ and $\delta E_K$ shown in Figs.~\ref{fig:AlftestdEratio} and \ref{fig:Alftest_coremaps} could be due to that this simulated cloud, being on the shock front of the collision of two diffuse clumps, is in principle trans-Alfv{\'e}nic.
While this implies that such equipartition may not be a general property for all simulations, it is very possible that in reality, all star-forming regions are indeed trans-Alfv{\'e}nic, and thus the DCF method could still provide a good approximation of the field strength measurement. 
This also explains why this equipartition tends to break when only including the immediate surrounding of a dense core, because the gas flows around a dense core are likely affected by the core's self-gravity, so the assumption of pure MHD turbulence no longer holds.

\subsection{The Statistical DCF Method}
\label{sec:statcorr}


The DCF method is based on the assumption of the equipartition between the perturbed components of gas velocity and magnetic field, i.e.,~$\delta E_K = \delta E_B$. 
As discussed in the previous section, 
one can use the vector-averaged field $\overline{\mathbf{B}}$ as the reference field and achieve the equipartition statistically in log space, i.e.,~$\langle\delta E_K/\delta E_B\rangle_{\rm log} \approx 0$, where we use $\langle\rangle_{\rm log}$ to represent averaged value in log space.
Therefore, the DCF equation should be re-written from Eq.~(\ref{eq:DCF3D}) as the following:
\begin{equation}
    B_{\rm DCF} = \left\langle B_{\rm DCF, local}\right\rangle_{\rm log}
    = \left\langle\sqrt{4\pi\rho}\frac{v_\perp}{ B_\perp/|\overline{\mathbf{B}}|}\right\rangle_{\rm log}.\label{eq:MDCF3D}
\end{equation}

Note that we propose here a slightly different process of applying the DCF method.
Instead of calculating 
\begin{equation}
    \langle B\rangle_{\rm DCF} \rightarrow \sqrt{4\pi\langle\rho\rangle}\frac{\langle v_\perp\rangle}{\langle B_\perp/|\overline{\mathbf{B}}|\rangle} 
    \label{eq:tradDCF}
\end{equation}
(see Sec.~\ref{sec:synPol} for more discussions on the DCF equation in 2D),
the log-normal shape of $\delta E_K/\delta E_B$ shown in Fig.~\ref{fig:AlftestdEratio} suggests that one should calculate the DCF-derived magnetic field strength at each location using local values of density, velocity, and $B_\perp/|\overline{\mathbf{B}}|$ as 
\begin{equation}
    B_{\rm DCF, local} = \sqrt{4\pi\rho} \frac{v_\perp} {B_\perp/|\overline{\mathbf{B}}|},
    \label{eq:BDCFlocal}
\end{equation}
then use $B_{\rm DCF} = \langle B_{\rm DCF, local}\rangle_{\rm log}$ as the cloud-scale magnetic field strength.\footnote{{Note that, in principle, there can be locations where $\mathbf{B}\parallel \overline{\mathbf{B}}$, i.e.,~$B_\perp = 0$. In this case, $B_{\rm DCF, local} \rightarrow \infty$. These data points should be removed before deriving the averaged value of the magnetic field. }}
Namely, while the DCF method cannot be applied locally at individual cells unless $\hat{\mathbf{B_0}}$ is known, we can take the vector-average of the magnetic field over a large-enough scale to approximate the unperturbed field direction, and use this direction to calculate $v_\perp$, $B_\perp$, and $B_{\rm DCF, local}$ at each cell. The averaged value of the distribution function of this $B_{\rm DCF, local}$ in log space can then be adopted as the estimated field strength of the system. 
{Note that, by comparing Eq.~(\ref{eq:MDCF3D}) with Eq.~(\ref{eq:DCF3D}), in Eq.~(\ref{eq:MDCF3D}) we are calculating $B_{\rm DCF}$ to approximate $|\overline{\mathbf{B}}|$, 
the vector-averaged field strength. 
However, the cloud-scale mean field strength should be $\overline{|\mathbf{B}|}$,
and obviously $|\overline{\mathbf{B}}| < \overline{|\mathbf{B}|}$.}
{As a result, using $B_\perp/|\overline{\mathbf{B}}|$ as the denominator in Eq.~(\ref{eq:MDCF3D}) tends to underestimate the field strength. }

{We further note that, a commonly adopted convention 
in previous 2D works \citep[e.g.,][]{FG08} is to consider 
\begin{equation}
    \tan\delta\phi \approx B_\perp / |\overline{\mathbf{B}}|
\end{equation}
where $\delta\phi = \angle(\mathbf{B},\overline{\mathbf{B}})$.
However, the correct relation should be $\tan\delta\phi = B_\perp/B_\parallel$ where $B_\parallel$ is the component of the local magnetic field parallel to $\overline{\mathbf{B}}$, and $B_\parallel\neq  |\overline{\mathbf{B}}|$. This suggests that by adopting $\tan\delta\phi$ in the DCF equation to replace $B_\perp/|\overline{\mathbf{B}}|$ would introduce additional errors. }
A better way is to consider $\sin\delta\phi = B_\perp/B$, i.e.,~using the local ratio between the perturbed (the component perpendicular to the mean field) and the total magnetic field to replace $B_\perp/|\overline{\mathbf{B}}|$ in Eq.~(\ref{eq:MDCF3D}).
{Note that similar concept has been pointed out in \cite{JHLiu_DCF_2021}.}
Though $B$ could have large variation within the interested region, since we are only considering the peak value in log space and since $\langle \delta E_K/\delta E_B \rangle_{\rm log} \approx 1$, Eq.~(\ref{eq:MDCF3D}) gives $B_{\rm DCF}\rightarrow \langle B\rangle_{\rm log} \approx \overline{|\mathbf{B}|}$.
Note that it does not matter if the derived $B_{\rm DCF, local}$ is locally correct or not, as we discussed above.
The right panel of Fig.~\ref{fig:AlftestBavg} compares the derived $B_{\rm DCF, local}$ values using $\tan\delta\phi = B_\perp/|\overline{\mathbf{B}}|$ (solid lines) and $\sin\delta\phi = B_\perp/B$ (dashed lines). The difference is small for locally-averaged field (small $n$) due to smaller $\delta\phi$, but for $B_{\rm DCF, local}$ derived from whole-box averaged field, switching to $\sin\delta\phi$ clearly shift the peak closer to the true value.
{As we will show in the next section, in projected 2D maps $\sin\delta\psi$ and $B_{{\rm POS},\perp} / B_{\rm POS}$ also have tighter correlation than $\tan\delta\psi$ vs.~$B_{{\rm POS},\perp} / \overline{B_{\rm POS}}$ (see Fig.~\ref{fig:Bvsp}).}


\section{The DCF Method in 2D: Complications and Modifications}
\label{sec:synobs}

Now that we have validated the DCF relation in 3D space, 
we extend our investigation of the DCF method to 2D projected systems, i.e.,~astronomical observations. 
We note that some extra assumptions are needed to apply the DCF method in observations due to the limited observables. Here, we discuss various ways to improve and/or validate these assumptions before applying the DCF method to synthetic observations from our numerical simulations in the next section.

\subsection{Synthetic Polarization}
\label{sec:synPol}

\begin{figure}
\begin{center}
\includegraphics[width=0.75\columnwidth]{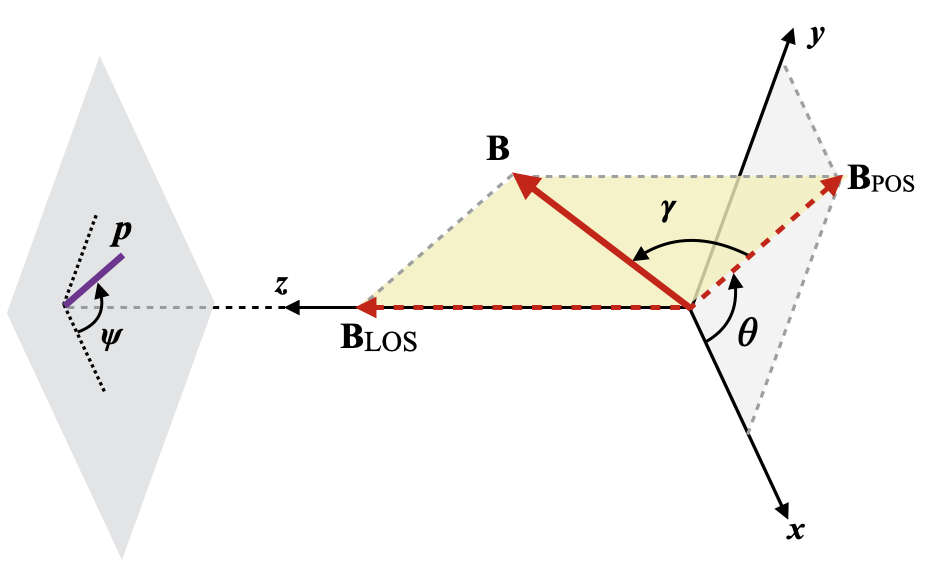}
\caption{\small Definition of angles and symbols in calculating the synthetic polarization. Revised from \protect\cite{Chen_HRO_2016, Chen2019}.}
\label{fig:angDef}
\end{center}
\end{figure}


We derived the synthetic polarization as the follows \citep[see e.g.,][]{Lee_Draine_1985,WK90,Fiege_Pudritz_2000,PlanckXIX,King_2018}:
\begin{align}
    N &= \sum_{\rm los} n,\ \ 
    N_2 = \sum_{\rm los} n\left(\cos^2\gamma - \frac{2}{3}\right),\notag \\
    q &= \sum_{\rm los} n\cos 2\theta\cos^2\gamma,\ \ 
    u =\sum_{\rm los} n\sin 2\theta\cos^2\gamma,\\
    \psi &= \frac{1}{2}{\rm arctan2}\left(u,q\right),\ \ 
    p = p_0 \frac{\sqrt{q^2+u^2}}{N - p_0 N_2}\notag
\end{align}
with $n$ being the density of the medium (see Fig.~\ref{fig:angDef} for definition of angles). 
Note that here we use $\theta$ for the position angle of the plane-of-sky component of the magnetic field ($\mathbf{B}_{\rm POS}$) instead of the commonly-adopted $\phi$ to avoid confusion with $\delta\phi$, which measures the angle between $\mathbf{B}_{\rm POS}$ and $\overline{\mathbf{B}_{\rm POS}}$. 
For simplicity, we assumed homogeneous grain alignment and set the polarization coefficient to be a constant, $p_0 = 0.1$ \citep[see e.g.,][]{Chen_HRO_2016,Chen2019}. However, we note that non-constant grain alignment efficiency may have effects on the polarization structure, as discussed in e.g.,~\cite{King19}.

\begin{figure}
\begin{center}
\includegraphics[width=\columnwidth]{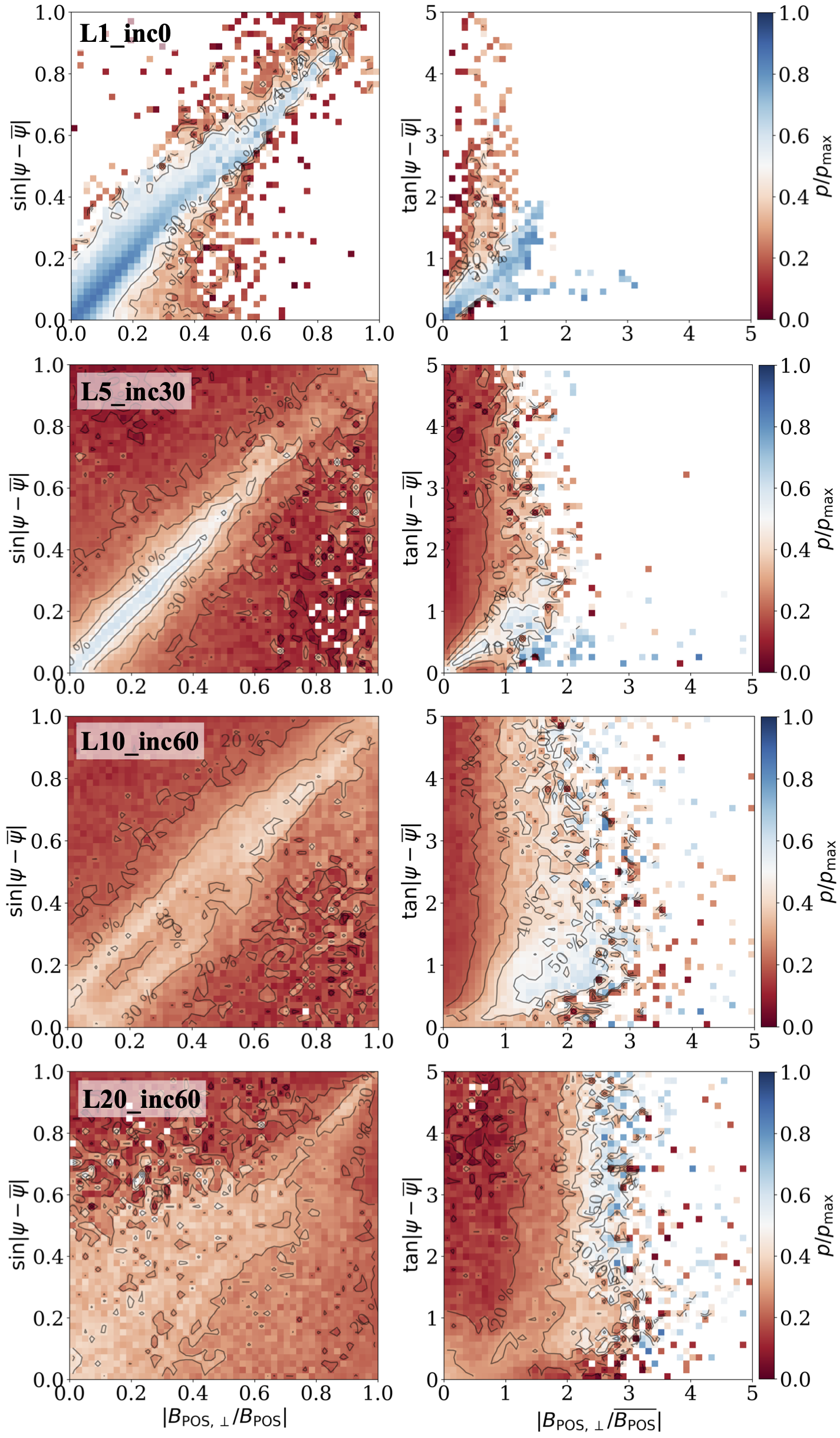}
\vspace{-.2in}
\caption{\small Comparison between the synthetic polarization structure (as traced by $\delta\psi = \psi-\overline{\psi}$ where $\overline{\psi}$ is the mean polarization angle over the entire map) and the projected plane-of-sky (POS) magnetic field structure {for 4 selected cases with various turbulent levels and viewing angles with respect to the magnetic field (more perturbed/inclined from top to bottom; see Table~\ref{tab:synobs}). As discussed in Sec.~\ref{sec:statcorr} and \ref{sec:synPol}, assuming $\tan\delta\psi\approx B_{\rm POS,\perp} / \overline{B_{\rm POS}}$ would increase the uncertainties in the DCF relation because $\tan\delta\psi$ does not trace $B_{\rm POS,\perp} / \overline{B_{\rm POS}}$ well (right column). Instead, $\sin\delta\psi$ and $B_{\rm POS,\perp}/B_{\rm POS}$ are more tightly correlated (left column).}
These 2D histograms are color-coded by normalized polarization fraction $p/p_{\rm max}$, which suggests that {in most of the cases} when the synthetic polarization fraction is relatively high, the synthetic polarization orientation is more consistent with the projected magnetic field structure. }
\label{fig:Bvsp}
\end{center}
\end{figure}

In addition to assuming isotropic turbulence so that $v_\perp \sim \sigma_v$ where $\sigma_v$ is the observed line-of-sight velocity dispersion,
to connect the 3D DCF relation to 2D projected quantities on the plane of sky (POS), we need to assume
\begin{equation}
{    
    \frac{B_\perp}{\overline{B}} \approx \frac{B_{\rm POS,\perp}}{\overline{B_{\rm POS}}} \ \ \ {\rm and} \ \ \  \frac{B_{\rm POS,\perp}}{\overline{B_{\rm POS}}}\approx \tan\delta\psi},
    \label{eq:DCFas}
\end{equation}
where $\delta\psi$ is the angle between $\psi$ and the averaged angle $\overline{\psi}$:
\begin{equation}
    \overline{\psi} \equiv \frac{1}{2}{\rm arctan2}\left(\sum_{\rm POS} \sin 2\psi, \sum_{\rm POS} \cos 2\psi\right).
\end{equation}
With these assumptions, Eq.~(\ref{eq:DCF3D}) becomes
\begin{equation}
    B_{\rm DCF, POS} \approx \sqrt{4\pi\rho}\frac{\sigma_v}{B_{\rm POS,\perp} / \overline{B_{\rm POS}}} \approx \sqrt{4\pi\rho}\frac{\sigma_v}{\tan\delta\psi}.
    \label{eq:DCF2Dlocal}
\end{equation}
Note that there have been several versions of the 2D DCF equation in the past, depending on the interpretation of the magnetic field distortion term $B_\perp/\overline{B}$. \cite{OSG2001} considered $\langle B_\perp/\overline{B}\rangle \sim\langle\delta\psi\rangle$, with $\langle\delta\psi\rangle$ being the dispersion of observed polarization angle. \cite{Heitsch01} used $B_\perp/\overline{B} \sim \langle \tan\delta\psi \rangle$, which could be severely contaminated by large angles. First proposed by \cite{FG08} and recently justified by \cite{PSLi_2021}, the now commonly adopted version of the DCF equation in 2D is 
\begin{equation}
    B_{\rm obs} = \langle B\rangle_{\rm DCF, POS} = \sqrt{4\pi\langle\rho\rangle}\frac{\langle\sigma_v\rangle}{\tan\langle\delta\psi\rangle}.
    \label{eq:DCF2D}
\end{equation}

{The accuracy of Eq.~(\ref{eq:DCF2D}) depends on how accurate Eq.~(\ref{eq:DCFas}) is.}
The first part of Eq.~(\ref{eq:DCFas}) {($B$ to $B_{\rm POS}$)} depends on the projection effect of the system, which we will discuss shortly in Sec.~\ref{sec:2Dproj}.
Regarding the second part of Eq.~(\ref{eq:DCFas}) {($B_{\rm POS}$ to polarization angle)},
a direct comparison between $B_{\rm POS,\perp}/B_{\rm POS}$ and $p_\perp/p = \sin(\psi-\overline{\psi})$, as well as $B_{\rm POS,\perp}/\overline{B_{\rm POS}}$ and $\tan(\psi-\overline{\psi})$, is shown in Fig.~\ref{fig:Bvsp} {using several synthetic observations (see Table~\ref{tab:synobs}).
As we discussed in Sec.~\ref{sec:statcorr}, $\sin\delta\psi$ correlates much better with $B_{\rm POS,\perp}/B_{\rm POS}$ compared with the correlation between $\tan\delta\psi$ and $B_{\rm POS,\perp}/\overline{B_{\rm POS}}$. This strengthen our argument in Sec.~\ref{sec:statcorr} that one should consider using $\sin\delta\psi$ in the DCF equation.}
Fig.~\ref{fig:Bvsp} {also} suggests that, {in non-extreme conditions (moderate viewing angle with respect to the magnetic field, moderate turbulence, etc.),} the polarization orientation follows the actual POS magnetic field structures pretty well when the polarization fraction is high enough, say, $p/p_{\rm max} \gtrsim 20\%$, where $p_{\rm max}$ is the maximum polarization fraction measured from the synthetic polarization map. Since the orientation of polarization segments with polarization fraction below $\sim 20\% p_{\rm max}$ have relatively low correlation with the actual magnetic field direction, we shall neglect those polarization segments when applying the DCF method to the observations. We discuss this in more details in Sec.~\ref{sec:DcutTest} below.

\subsection{Angle Correction: 2D Projection of 3D Angle}
\label{sec:2Dproj}

\begin{figure}
\begin{center}
\includegraphics[width=0.45\textwidth]{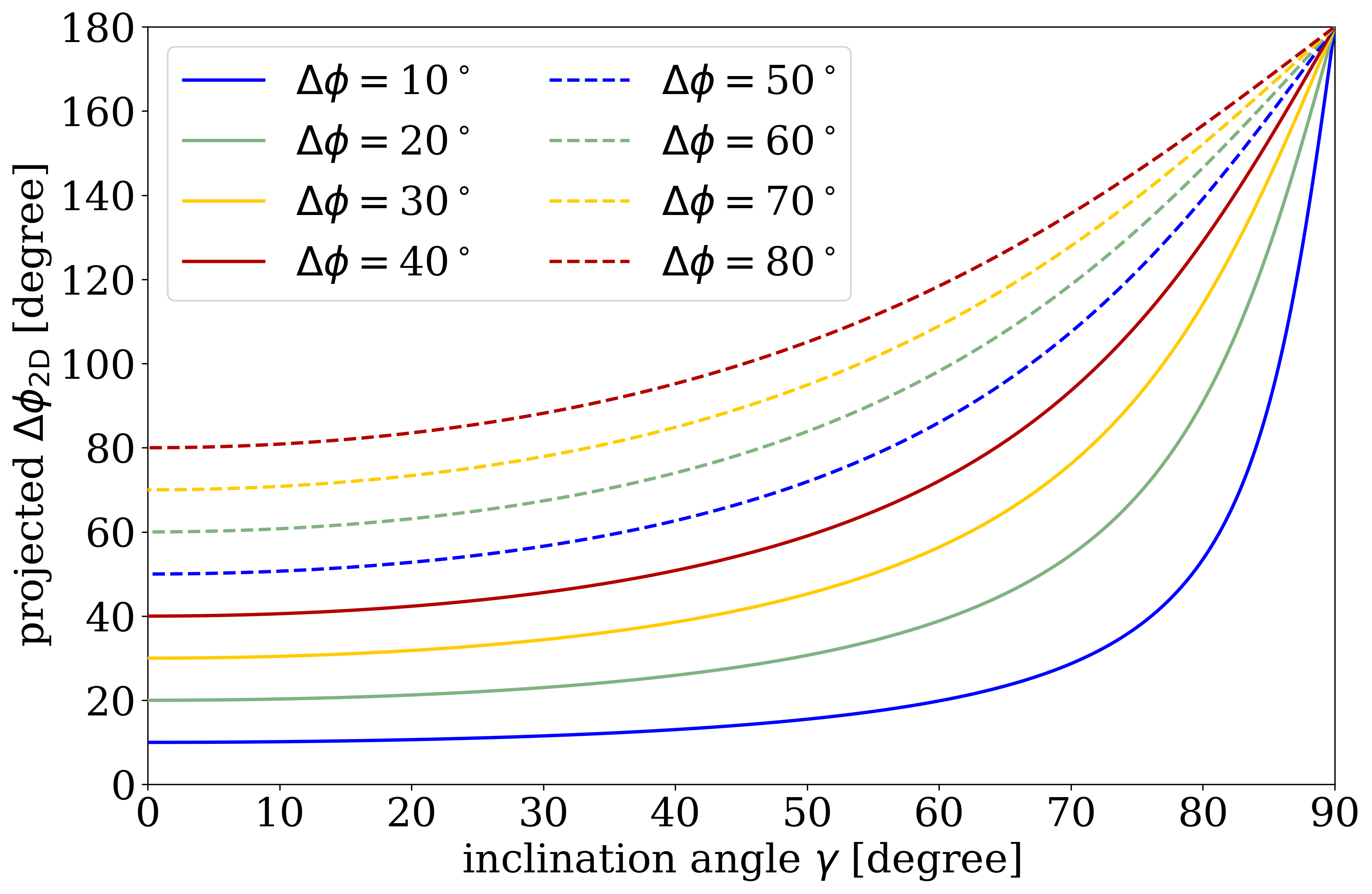}
\caption{\small The projected angle {$\Delta\phi_{\rm 2D}$} from various values of {$\Delta\phi$}, as functions of the angle of the projection $\gamma$ (i.e.,~the inclination angle of the plane of sky).}
\label{fig:ProjAng}
\end{center}
\end{figure}

As discussed above, the DCF-derived magnetic field strength depends on the angle between the perturbed component and the mean direction of the magnetic field. 
However, it is important to note that the angle between two vectors in 3D may not be the same after being projected to 2D plane, and thus the uncertainty increases when applying the DCF method to 2D projected observations.
Here we derive the equation for estimating the projection effect from 3D to 2D; i.e.,~the relation between the projected angle in 2D and the actual angle in 3D.

Consider two unit vectors on 3D Cartesian coordinates with an angle $\alpha$ between them. Without loss of generality, we set these two vectors to be on the $x-y$ plane with
\begin{equation}
    \mathbf{r}_\pm = \cos \frac{\alpha}{2} \hat{\mathbf{i}} \pm \sin\frac{\alpha}{2}\hat{\mathbf{j}}.
\end{equation}
After projecting these two vectors to a plane ${\cal H}$ with normal vector $\hat{\mathbf{d}} = \sin\Theta\cos\Phi\hat{\mathbf{i}} + \sin\Theta\sin\Phi\hat{\mathbf{j}}+\cos\Theta\hat{\mathbf{k}}$, the projected vectors become
\begin{equation}
    \mathbf{p}_\pm = \mathbf{r}_\pm - (\mathbf{r}_\pm \cdot \hat{\mathbf{d}})\hat{\mathbf{d}}.
\end{equation}
Let the angle between the projected vectors be $\alpha_{\rm proj}$, then we have
\begin{align}
    \cos\alpha_{\rm proj} &= \frac{\mathbf{p_+}\cdot \mathbf{p_-}}{|\mathbf{p_+}| |\mathbf{p_-}|} \notag \\
    &=\frac{\cos\alpha - (\mathbf{r_+}\cdot\hat{\mathbf{d}})(\mathbf{r_-}\cdot\hat{\mathbf{d}})}{\sqrt{1-(\mathbf{r_+}\cdot\hat{\mathbf{d}})^2}\sqrt{1-(\mathbf{r_-}\cdot\hat{\mathbf{d}})^2}}.\label{eq:projBeta}
\end{align}
Since
\begin{equation}
        \mathbf{r}_\pm\cdot\hat{\mathbf{d}} = \sin\Theta\cos\left(\Phi\mp\frac{\alpha}{2}\right),
\end{equation}
we have
\begin{equation}
    (\mathbf{r_+}\cdot\hat{\mathbf{d}})(\mathbf{r_-}\cdot\hat{\mathbf{d}}) = \frac{1}{2}\sin^2\Theta\left[\cos(2\Phi)+\cos\alpha\right],
\end{equation}
and
\begin{align}
    (\mathbf{r}_\pm\cdot\hat{\mathbf{d}})^2 &= \sin^2\Theta\cos^2\left(\Phi\mp\frac{\alpha}{2}\right)\notag \\
    & = \frac{1}{2}\sin^2\Theta\left[\cos(2\Phi\mp\alpha)+1\right].
\end{align}

For projected magnetic field on the plane of sky, $\mathbf{r}_\pm$ represent two 3D vectors with angle {$\Delta\phi$} between them, and $\mathbf{p}_\pm$ are two vectors on the 2D map with angle difference {$\Delta\phi_{\rm 2D}$}. If we use the average of the two vectors as the reference direction and let the inclination angle between the mean field direction and the plane of sky be $\gamma$, the normal vector for ${\cal H}$ (the plane of sky) is on the $x-z$ plane with $\Phi = 0$ and $\Theta = \pi-\gamma$. Note that since we only consider $\gamma$ values within $[0,\pi/2]$, this means $\Theta$ must be within $[\pi/2,\pi]$. Eq.~(\ref{eq:projBeta}) thus becomes:
\begin{equation}
    \cos\Delta\phi_{\rm 2D} = \frac{\cos\Delta\phi - \frac{1}{2}\sin^2\gamma(1+\cos\Delta\phi)}{1-\frac{1}{2}\sin^2\gamma(1+\cos\Delta\phi)},
    \label{eq:3DProj}
\end{equation}
and we now have the correlation between projected angle {$\Delta\phi_{\rm 2D}$} and the original angle $\Delta\phi$ with $\gamma$ being the inclination angle of the mean magnetic field with respect to the plane of sky. 

Fig.~\ref{fig:ProjAng} illustrates this projection effect by showing {$\Delta\phi_{\rm 2D}$} as functions of both {$\Delta\phi$} and $\gamma$. Note that the projected angle measured in 2D is always larger than the actual angle in 3D, and the difference increases with inclination angle. This suggests that $B_\perp/B$ tends to be slightly overestimated, which makes the DCF-derived magnetic field slightly underestimated. However, this effect could be small if the inclination angle is not large (small $\gamma$) or the actual perturbed field does not deviate much from the mean field in 3D (small {$\Delta\phi$}). For example, from Fig.~\ref{fig:ProjAng}, for a dispersion angle $\delta\phi=20^\circ$ in 3D, the DCF method shall remain accurate within a factor of two if the inclination angle is roughly below $60^\circ$. 
Nevertheless, as shown in Sec.~\ref{sec::test}, this projection correction is critical particularly in cases with large inclination angles (see e.g.,~ Fig.~\ref{fig:synComp1}).

\subsection{Gas Volume Density in the DCF Method}
\label{sec:rho}


When applying the DCF method on observations, one challenge is the estimate of the gas volume density $\rho$, which requires additional information either from a chemical network or a measurement of the cloud depth along the line of sight. Here we describe two methods on estimating the cloud depth and hence the gas volume density from observable quantities.  We compare them with the actual values and discuss how this affects the accuracy of the DCF method in Sec.~\ref{sec::compModel}.


\begin{figure}
\begin{center}
\includegraphics[width=0.9\columnwidth]{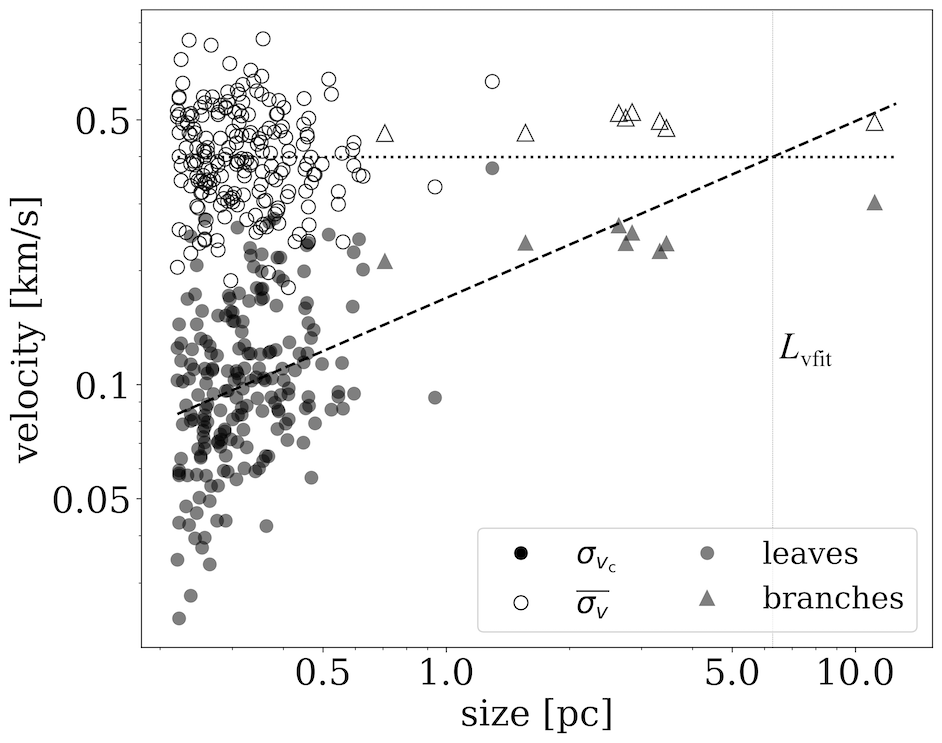}
\vspace{-.1in}
\caption{\small The linewidth-size correlation derived from dendrogram structures in model L10\_inc30. {The fitted cloud depth $L_{\rm vfit}$ (thin gray vertical line) is considered as the intersection of the mean line-of-sight velocity dispersion $\overline{\sigma_v}$ (dotted horizontal line) and the fitted linewidth-size correlation (dashed line).} }
\label{fig:synLW}
\end{center}
\end{figure}

\subsubsection{Linewidth-Size Correlation}
\label{sec:vfit}

We first consider the method proposed in \cite{Storm14} to estimate the cloud depth.
This method (`vfit') is based on the assumption that the observed spectral linewidth traces the velocity dispersion ($\sigma_v$) corresponding to the line-of-sight length scale of the cloud, while the spatial dispersion of the observed centroid velocity ($v_c$) should reflect the plane-of-sky size of the cloud based on Larson's law on linewidth-size correlation (see Fig.~\ref{fig:synLW} for illustration). 
{We note that this assumption is valid as long as the entire target region belongs to a spatially-connected structure with a power-law correlation between the turbulence amplitude and the physical scale. Such linewidth-size correlation may break down in clouds with significant sub-structures, however. }

Following \cite{Storm14}, 
we identify spatially-coherent structures (cores, branches, trunks) from the column density map using the \texttt{Python} toolkit \texttt{astrodendro},
then calculate the mean linewidth ($\overline{\sigma_v}$) and the dispersion of centroid velocity ($\sigma_{v_c}$) within each structure. We then fit the dispersion of centroid velocity as a power-law function of the physical size of the structure, $\ell$:
\begin{equation}
    {\sigma_{v_c}} = v_1\ell^\alpha,\label{eq:vfit}
\end{equation}
which represents the scale relation of the turbulence of the given system, $\sigma_{v_c}(\ell)$. 
We can then use the average value of the line-of-sight velocity dispersion to determine the depth of the cloud $L$ {under the assumption that the turbulence is isotropic}:
\begin{equation}
    \overline{\sigma_v} = {\sigma_{v_c}}(\ell = L_{\rm vfit}) = v_1 L_{\rm vfit}^\alpha,\ \ \ 
    L_{\rm vfit} = \left(\frac{\overline{\sigma_v}}{v_1}\right)^{1/\alpha}.
    \label{eq:vfitL}
\end{equation}

Note that, when applying on the DCF analysis, this method is relatively independent of the line-of-sight velocity dispersion if the scale dependence of the centroid velocity roughly follows the Larson's law, i.e.,~$\alpha\approx 0.5$:
\begin{equation}
    B_{\rm vfit} = \sqrt{4\pi\rho}\frac{\sigma_v}{\tan\delta\psi} = \sqrt{4\pi\frac{\Sigma}{L_{\rm vfit}}}\frac{\sigma_v}{\tan\delta\psi} \approx 
    \frac{\sqrt{4\pi\Sigma\cdot v_1^2}}{\tan\delta\psi}.
\end{equation}
The derived magnetic field strength thus only depends on the fitting result of $v_1$, the dispersion of centroid velocity within unit-size structures. 

\subsubsection{Pressure Equilibrium}
\label{sec:eq}

Alternatively, we can consider the cloud as a layer in hydrostatic equilibrium \citep[see e.g.,][]{EE78}, but in a more general form. 
Considering a self-gravitating sheet-like cloud where the internal energy density balances the pressure from gravitational potential:
\begin{equation*}
    {\cal E}_{\rm int} = {\cal E}_{\rm grav},
\end{equation*}
where ${\cal E}_{\rm grav} = \rho \Phi_{\rm grav}$ and $\nabla^2\Phi_{\rm grav} = -4\pi G\rho$.
For a magnetized turbulent cloud, we shall have ${\cal E}_{\rm int} = {\cal E}_{\rm thermal} + {\cal E}_{\rm turb} + {\cal E}_B$. Assuming constant density, the equation can be simplified as
\begin{equation}
    \frac{1}{2}\rho (c_s^2 + \sigma_v^2)+ \frac{B^2}{8\pi} = \rho\Phi_{\rm grav} = \frac{\pi G \Sigma^2}{2},
    \label{eq:pres_eq}
\end{equation}
where $\pi G\Sigma^2/2$ is the gravitational pressure in the midplane \citep[see e.g.,][]{McKee93}.

Combining with the 2D DCF equation Eq.~(\ref{eq:DCF2D}), we can solve the cloud depth $L = \Sigma/\rho$ as
\begin{equation}
    L_{\rm eq} = \frac{\sigma_v^2\left((B_\perp/B)^{-2} + 1\right) + c_s^2}{\pi G \Sigma}.
    \label{eq:Leq}
\end{equation}
The corresponding magnetic field strength is therefore
\begin{equation}
    B_{\rm eq} = \sqrt{4\pi\frac{\Sigma}{L_{\rm eq}}}\frac{\sigma_v}{\tan\delta\psi} = \left[ \frac{4\pi^2 G \Sigma^2}{\tan^2\delta\psi\left(1+\left(\frac{c_s}{\sigma_v})\right)^2\right) + 1} \right]^{1/2}.
    \label{eq:Bslab}
\end{equation}
Note that $\tan\delta\psi \approx B_\perp/\overline{B}$ is basically $\sqrt{4\pi\rho}\sigma_v/\overline{B} = {\cal M}_A$, the Alfv{\'e}n Mach number. In the case of magnetic domination over turbulence, ${\cal M}_A \ll 1$, Eq.~(\ref{eq:Bslab}) reduces to
\begin{equation*}
    \overline{B} \approx 2\pi\sqrt{G}\Sigma,
\end{equation*}
which means the normalized mass-to-flux ratio of the cloud $\Sigma/\overline{B} \cdot 2\pi\sqrt{G} \approx 1$, or the cloud is magnetically critical. {We stress that the method described in this subsection is applicable only to self-gravitating sheet-like clouds.}

\begin{figure*}
\begin{center}
\includegraphics[width=\textwidth]{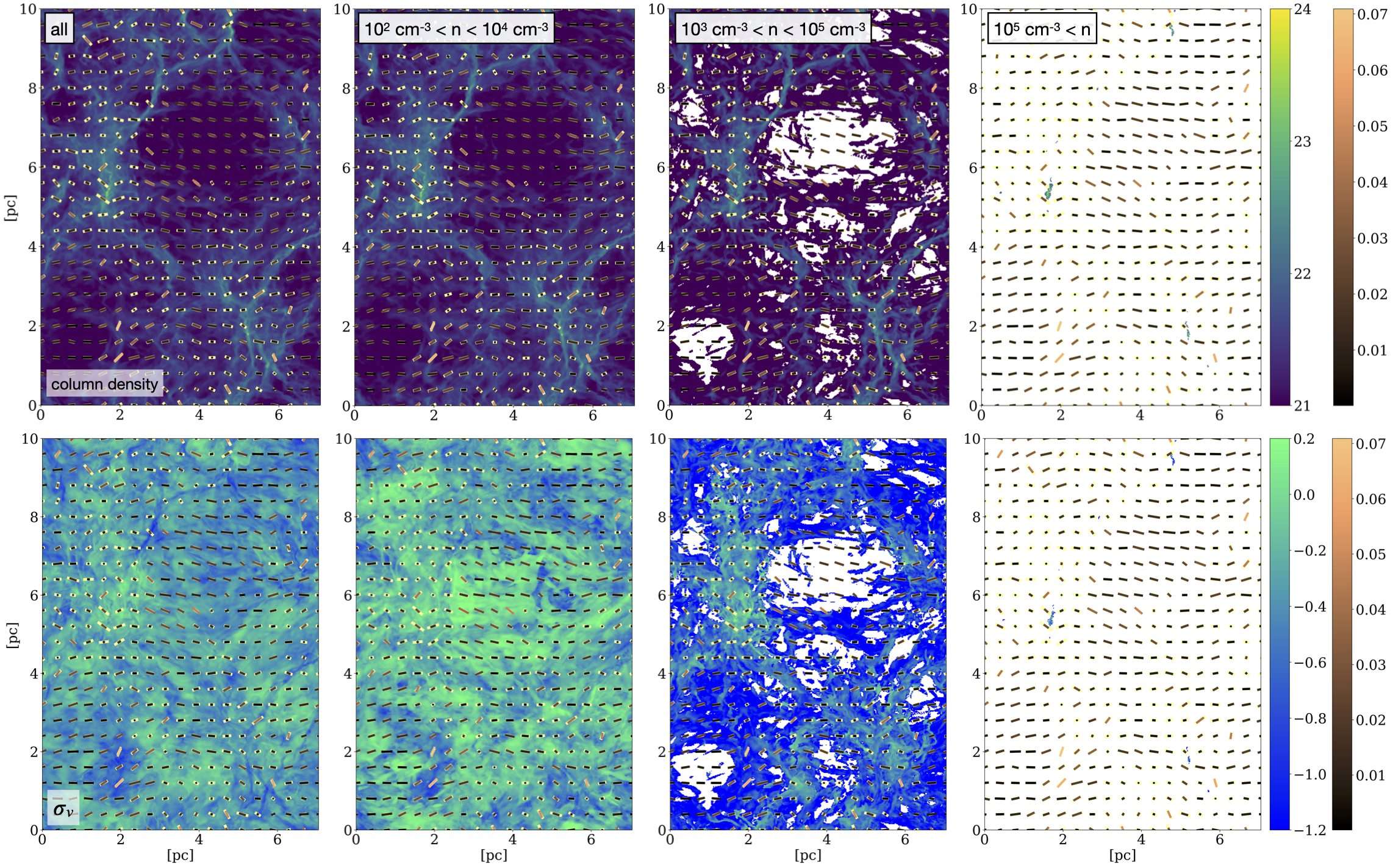}
\vspace{-.2in}
\caption{\small Demonstrating the density cutoff test, showing the corresponding integrated column density (in log scale with polarization segments color-coded by polarization fraction; {\it top}) and the line-of-sight velocity dispersion (in log scale; {\it bottom}). 
The case of $10^2 < n/[{\rm cm}^{-3}] < 10^4$ ({\it second column}) seems to better resemble the `all' case (i.e.,~including all cells and all density ranges; {\it left column}),
while the extreme case $10^5 < n/[{\rm cm}^{-3}]$ ({\it right column}) only traces the densest structures and has too few cells to conduct meaningful analyses.  }
\label{fig:Dcut}
\end{center}
\end{figure*}

\subsection{Density Cutoffs as Synthetic Line Observations}
\label{sec::SynLine}

In addition to gas volume density, 
the DCF method still requires more information than just the polarization morphology.
Observationally, the velocity dispersion $\sigma_v$ can be probed by the width of molecular line emission in the velocity space. However, the critical gas densities of various molecular lines are different, and the measured velocity dispersion therefore may highlight different regions of the target cloud that have different densities. Synthetic line observations using tracers with different critical densities may help test how this could affect the accuracy of the DCF method.

We thus consider a simplified method of generating synthetic line observations by applying density masks to all sightlines; i.e.,~for each sightline, we only include cells with gas volume densities within the specified density range when generating the corresponding synthetic observations.
This is illustrated in Fig.~\ref{fig:Dcut} using model L10\_inc30 as an example. Three density ranges are selected to approximate the typical gas tracers commonly used in line observations towards star-forming regions \citep[see e.g.,][]{Shirley_critrho_2015, Fissel_2019}: $10^2< n/[{\rm cm}^{-3}]<10^4$ (low to intermediate density regime, e.g.,~$^{13}$CO, C$^{18}$O), $10^3<n/[{\rm cm}^{-3}]< 10^5$ (intermediate to dense gas, e.g.,~NH$_3$, N$_2$H$^+$), and $10^5< n/[{\rm cm}^{-3}]$ (the densest gas component, e.g.,~HCN, H$^{13}$CO$^+$).
Note that we only used our simplified synthetic line observation to calculate the synthetic integrated line intensity (used in defining dendrograms for the `vfit' density-estimating method) and velocity dispersion, but not the column density (which is typically derived from multi-wavelength continuum observations) and polarization angles, because thermal dust emission is independent from the molecular line observations.

Fig.~\ref{fig:Dcut} demonstrate this synthetic line observation test by showing the corresponding integrated emission (top row) and velocity dispersion (bottow row) from the three cases described above. 
Obviously, dense gas tracers ($10^5< n/[{\rm cm}^{-3}]$) would only follow the densest structures, and thus is not very suitable for our statistical approach of the DCF method. 
Also, we note that the velocity dispersion traced by the low-density tracer ($10^2< n/[{\rm cm}^{-3}]<10^4$) is higher than that measured in gas with intermediate densities ($10^3< n/[{\rm cm}^{-3}]<10^5$), which is not surprising because we expect the diffuse gas to be more turbulent. 
We will discuss the corresponding DCF results in Sec.~\ref{sec:DcutTest} below.

\section{Comparisons and Discussions}
\label{sec::test}

\subsection{Testing density-estimating methods in synthetic observations with the statistical approach}
\label{sec::compModel}

\begin{figure*}
\begin{center}
\includegraphics[width=\textwidth]{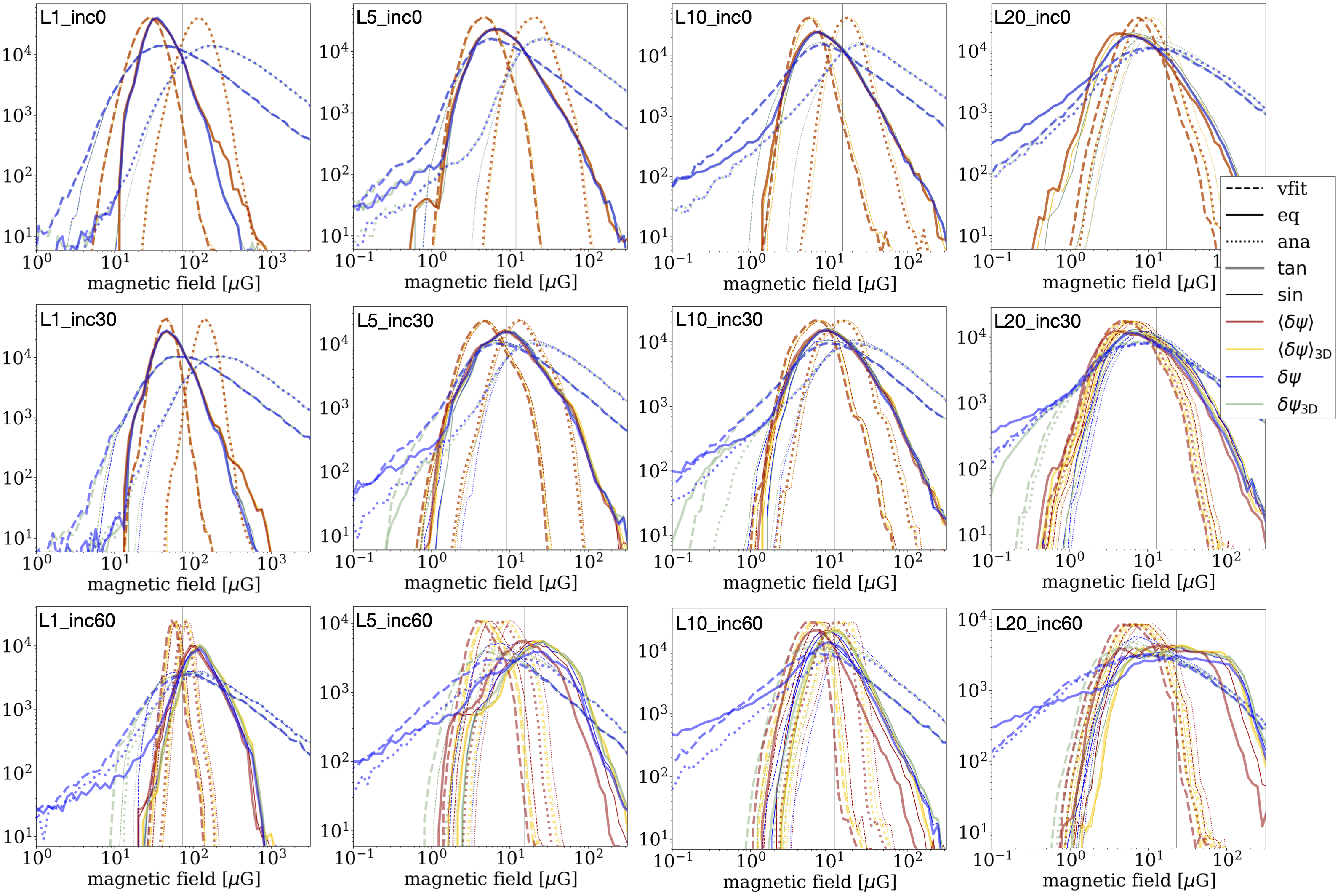}
\vspace{-.2in}
\caption{\small The PDFs of DCF-derived magnetic field strength using Eq.~(\ref{eq:DCF2Dlocal}) for {all} synthetic observations {considered in this study, with 4 simulation models (each column; see Table~\ref{tab:synobs}) and 3 viewing angles (each row)}, roughly following the order of relative turbulent strength in the plane of sky from left to right panels. In each panel, we compare the DCF-derived magnetic field strength from the three density-estimating methods discussed in Sec.~\ref{sec:rho} (dashed, solid, and dotted lines represent method `vfit', `eq', and `ana', respectively), as well as the potential modifications to the DCF method that we discussed earlier: switching $\langle\delta\psi\rangle$ with $\delta\psi$ (red/yellow vs.~blue/green curves), using $\sin\delta\psi$ instead of $\tan\delta\psi$ (thick vs.~thin lines), and considering the projection effect to estimate the dispersion angle in 3D (yellow and green curves). See text for more discussions. }
\label{fig:sample}
\end{center}
\end{figure*}

\begin{table*}
 \begin{center}
  \caption{Summary of the synthetic observations described in Sec.~\ref{sec::compModel} and the corresponding results from the DCF analysis with various modifications discussed in Sec.~\ref{sec:synobs}.}
  \label{tab:synobs}
  \begin{tabular}{lcccccccccccccccc}
    \hline
    \hline
    model & $\gamma_{\overline{\bf B}}$ & $\langle\delta\psi\rangle$ & $\overline{\sigma_v}$ & $\Sigma$ & \multicolumn{3}{c}{$\log (\overline{\rho}/{\rm cm}^{-3}$)} & \multicolumn{3}{c}{$B_{\rm DCF}/B_{\rm POS}$, $\tan\langle\delta\psi\rangle$} & \multicolumn{3}{c}{$B_{\rm DCF}/B_{\rm POS}$, $\tan\delta\psi$} & \multicolumn{3}{c}{$B_{\rm DCF}/B_{\rm 3D}$, $\sin\delta\psi_{\rm 3D}$}\\
    & $(^\circ)$ & $(^\circ)$ & (km/s) & ($10^{21}$\,cm$^{-2}$) & vfit & eq
    & ana & vfit & eq & ana & vfit & eq & ana & vfit & eq & ana \\
    \hline
    L1\_inc0 & \ 7 & \ 8 & 0.18 & 6.54 & 3.1 & 2.6 & 4.3 & 0.4 & 0.5 & 1.6 & 0.9 & 0.5 & 3.9 & 0.9 & 0.5 & 3.8\\
    L1\_inc30 & 37 & 11 & 0.28 & 8.80 & 3.3 & 2.7 & 4.3 & 0.8 & 0.9 & 2.4 & 1.8 & 0.9 & 5.8 & 1.5 & 0.7 & 4.7\\
    L1\_inc60 & 67 & 27 & 0.41 & 24.0 & 4.0 & 4.0 & 4.3 & 1.5 & 3.2 & 2.2 & 3.3 & 3.0 & 4.7 & 1.8 & 1.8 & 2.5\\
    \hline
    L5\_inc0 & \ 7 & 17 & 0.31 & 1.41 & 1.7 & 1.5 & 2.9 & 0.4 & 0.7 & 1.8 & 0.9 & 0.7 & 3.9 & 0.9 & 0.7 & 3.8\\
    L5\_inc30 & 37 & 24 & 0.31 & 1.85 & 2.0 & 1.9 & 2.9 & 0.6 & 1.2 & 1.8 & 1.5 & 1.2 & 4.2 & 1.3 & 1.1 & 3.7\\
    L5\_inc60 & 67 & 43 & 0.42 & 5.12 & 2.4 & 3.2 & 2.9 & 0.5 & 1.5 & 0.9 & 0.8 & 1.2 & 1.5 & 0.7 & 1.7 & 1.3\\
    \hline
    L10\_inc0 & \ 2 & 24 & 0.51 & 1.71 & 1.8 & 1.6 & 2.7 & 0.6 & 0.6 & 1.2 & 1.2 & 0.6 & 2.5 & 1.3 & 0.6 & 2.7\\
    L10\_inc30 & 28 & 28 & 0.49 & 2.22 & 2.1 & 1.9 & 2.7 & 0.6 & 1.0 & 1.4 & 1.4 & 0.9 & 2.9 & 1.4 & 1.0 & 2.8\\
    L10\_inc60 & 62 & 45 & 0.75 & 2.53 & 2.1 & 2.2 & 2.8 & 0.8 & 0.9 & 1.5 & 1.2 & 0.7 & 2.2 & 1.2 & 1.1 & 2.1\\
    \hline
    L20\_inc0 & \ 2 & 37 & 0.68 & 1.68 & 2.2 & 1.7 & 2.4 & 0.5 & 0.4 & 0.6 & 0.8 & 0.3 & 1.1 & 1.1 & 0.5 & 1.4\\
    L20\_inc30 & 32 & 42 & 0.56 & 2.13 & 2.3 & 2.2 & 2.4 & 0.5 & 0.6 & 0.6 & 0.9 & 0.5 & 1.0 & 1.1 & 0.8 & 1.2\\
    L20\_inc60 & 63 & 47 & 0.75 & 5.17 & 2.2 & 2.8 & 2.3 & 0.3 & 0.8 & 0.4 & 0.5 & 0.6 & 0.6 & 0.6 & 1.1 & 0.7\\
    \hline
    &&&& \multicolumn{4}{r}{$\epsilon\equiv \langle|\log (B_{\rm DCF}/B)|\rangle$} & 0.27 & 0.18 & 0.24 & 0.16 & 0.20 & 0.41 & 0.12 & 0.15 & 0.38\\
    \hline
  \end{tabular}
  \end{center}
\end{table*}

\begin{figure*}
\begin{center}
\includegraphics[width=\textwidth]{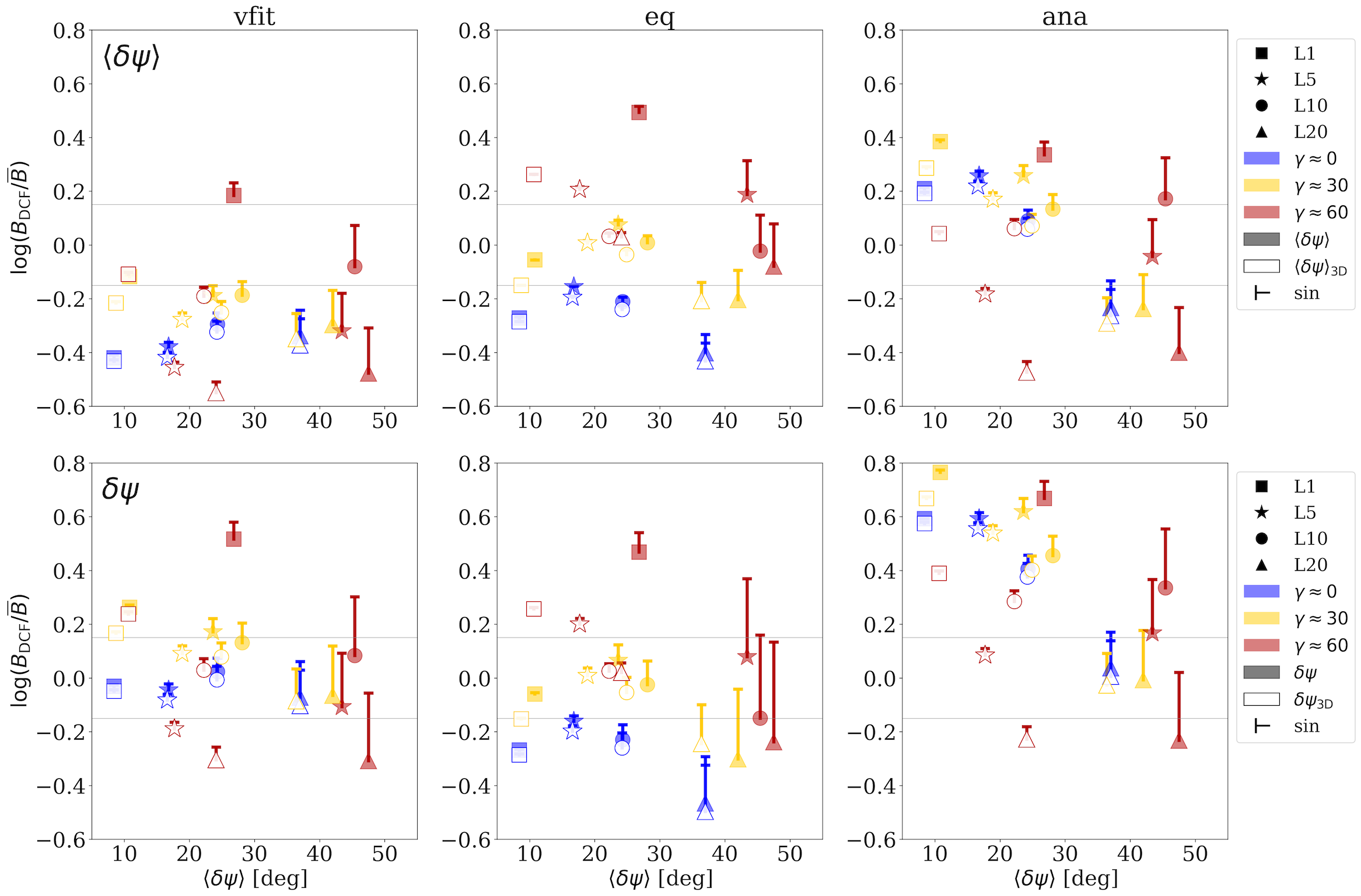}
\vspace{-.2in}
\caption{\small Summary of the analysis discussed in Sec.~\ref{sec::compModel}, plotting the ratio between the DCF-derived and the actual magnetic field strengths, in log scale, as functions of the mean dispersion angle $\langle\delta\psi\rangle$. Synthetic observations were generated from 4 different simulations ({\it different marker styles}) and 3 inclination angles of the mean magnetic field $\gamma$ ({\it different marker colors}). Three methods were applied to estimate the gas density from observables ({\it left, middle, and right columns}) as discussed in Sec.~\ref{sec:rho}. We also compared the traditional way of calculating $B_{\rm DCF}$ using $\tan\langle\delta\psi\rangle$ ({\it top row}) with our proposed revision, $\tan\delta\psi$ ({\it bottom row}). Also shown is the projection correction ({\it open symbols}) that convert the 2D angles to 3D, as discussed in Sec.~\ref{sec:2Dproj}. Another proposed revision that replaces $\tan\delta\psi$ with $\sin\delta\psi$ is shown as single-sided errorbars. We see clear improvement of the accuracy of $B_{\rm DCF}$ when switching from $\tan\langle\delta\psi\rangle$ to $\tan\delta\psi$, and the projection correction could be critical for cases with large inclination angles of the magnetic field ($\gamma\approx 60^\circ$; {\it red symbols}). In some cases with high inclination angles and/or large dispersion of polarization angles, switching to $\sin$ instead of using $\tan$ also increases the accuracy of $B_{\rm DCF}$. More importantly, while the results derived from method `ana' ({\it right column})) are supposed to be the most accurate, we found that `vfit' ({\it left column}) and `eq' ({\it middle column}) methods in general give better estimates of the actual mean field ($\overline{B_{\rm POS}}$ or $\overline{B}$), mostly within a factor of 2 (dotted gray horizontal lines mark $\pm\log 2$). }
\vspace{-.1in}
\label{fig:synComp1}
\end{center}
\end{figure*}

We first focus on comparing the different methods of deriving gas volume density using synthetic observations generated from our 3D MHD simulations.
Here, we consider 4 simulation models (L1, L5, L10, and L20; see Table~\ref{tab:models}) and generate synthetic observations at 3 viewing angles so that the inclination angles of the mean magnetic field are $\gamma\approx 0^\circ$, $30^\circ$, $60^\circ$. 
The basic observable properties of these synthetic observations that are important in the DCF analysis (column density, velocity dispersion, and polarization angle dispersion) are listed in Table~\ref{tab:synobs}.

The derived mean densities are also listed in Table~\ref{tab:synobs}. 
Besides the two density-estimating methods discussed in Sec.~\ref{sec:rho} (`vfit' and `eq'), 
we also provide estimates of the cloud depths and corresponding gas density $\rho_{\rm ana} = \Sigma_{\rm obs} / L_{\rm ana}$ from our 3D simulation data.
Since all models considered here are convergent flow simulations where clouds formed by shock compression, the typical depth of the cloud when viewed face-on is the thickness of the post-shock layer, which is about $10\%$ of the size of the simulation box. For simplicity, we define the `analytic' cloud depth at various inclination angle as
\begin{equation}
    L_{\rm ana} \equiv \frac{0.1 L_{\rm box}}{\cos\gamma}.
\end{equation}

In addition, three more comparisons are considered here: 
1) whether or not the previously proposed $\langle\delta\psi\rangle\rightarrow\delta\psi$ replacement based on our 3D analysis (see Sec.~\ref{sec:statcorr}) still holds in 2D observations,\footnote{Note that this is different from $\langle B\rangle_{\rm DCF}$ given in Eq.~(\ref{eq:DCF2D}), which uses the mean density and velocity dispersion to give one value of the DCF-derived field strength over the entire map. Here, we want to focus on the comparison between $\langle\delta\psi\rangle$ and $\delta\psi$, and thus we keep the local values of $\rho$ and $\sigma_v$ in the calculation and only switch between $\langle\delta\psi\rangle$ and $\delta\psi$ (see Eq.~(\ref{eq:DCF2Dlocal})). The shape of the distribution of DCF-derived field strength when using $\langle\delta\psi\rangle$ therefore completely depends on hydrodynamic properties of the gas (see Fig.~\ref{fig:sample}). }
2) whether or not the previously proposed $\tan\delta\psi\rightarrow\sin\delta\psi$ replacement based on our 3D analysis (see Sec.~\ref{sec:statcorr}) still holds in 2D observations,
and 3) if the projection correction on 2D dispersion angle could be critical in the DCF analysis.
A summary of the synthetic models and corresponding DCF results,
including
the mean densities derived from the linewidth-size correlation (`vfit'), the equilibrium (`eq'), and the analytic solution (`ana'), 
the DCF-derived magnetic field strength
using both $\langle\delta\psi\rangle$ and $\delta\psi$ as the polarization angle dispersion, as well as a projection-corrected DCF result using $\sin\delta\psi_{\rm proj}$,
can be found in Table~\ref{tab:synobs}, while Figs.~\ref{fig:sample} and \ref{fig:synComp1} provide more detailed comparisons of the results through graphical visualization.


Fig.~\ref{fig:sample} illustrates our results from the 12 synthetic observations, showing the probability distribution functions of the DCF-derived magnetic field strength 1) using different density-deriving methods, 2) with or without the replacement of $\langle\delta\psi\rangle\rightarrow\delta\psi$ or $\tan\delta\psi\rightarrow\sin\delta\psi$, and 3) with or without the projection correction $\delta\psi\rightarrow\delta\psi_{\rm 3D}$, which can be derived from Eq.~(\ref{eq:3DProj}):
\begin{equation}
    \delta\psi_{\rm 3D} = \delta\phi = \arccos{ \frac{\cos\delta\psi + \frac{1}{2}\sin^2\gamma\left(1-\cos\delta\psi\right)}{1-\frac{1}{2}\sin^2\gamma\left(1-\cos\delta\psi\right)}}.
    \label{eq:proj3D}
\end{equation}
Note that when considering the projection effect of $\delta\psi$, we adopted the projection-corrected velocity dispersion $\sigma_{v,{\rm 3D}} \equiv \sigma_v \cos\gamma$ to replace $\sigma_v$ in the DCF equation (Eq.~(\ref{eq:DCF2D})). This is to exclude the velocity component parallel to the magnetic field, which does not contribute to the observed field distortion.

Comparing Fig.~\ref{fig:sample} to the 3D results (right panel of Fig.~\ref{fig:AlftestBavg}; also see right panel of Fig.~\ref{fig:AlftestdEratio}), we see that the 2D DCF-derived field strengths do not always preserve the nice, symmetric shape of log-normal distribution of the DCF-derived field strengths in 3D. Only synthetic observations with moderate turbulence levels (e.g.,~models L5 and L10) and inclination angles (e.g.,~$\gamma\lesssim 30^\circ$) seem to recover the log-normal distribution of magnetic field strength, and in fact have more accurate DCF results (also see Table~\ref{tab:synobs}). This indicates that the DCF method is less applicable for extreme scenarios with strong turbulence {(e.g.,~model L20)} and/or very large inclination angles {(e.g.,~$\gamma \gtrsim 60$)}. This is not surprising, because the main assumptions one needs to make in order to apply the DCF method in 2D (isotropic velocity field and no cancellation along the line of sight) break in these extreme cases. 

Another important feature revealed by Fig.~\ref{fig:sample} is the clear difference in both the value and the shape of the distribution between the three density-deriving methods.
We also note, from Table~\ref{tab:synobs}, that the derived mean densities from the three different methods are highly inconsistent with each other in many of the synthetic observations we studied here.
This in fact provides an important message on applying the DCF method in observations: the biggest uncertainty of calculating DCF-derived magnetic field strength comes from the estimate of gas density, not the polarization measurement.
We will discuss this later in this section.

The complete results combining all 12 synthetic observations, 3 density-deriving methods, and 3 proposed modifications ($\langle\delta\psi\rangle\rightarrow\delta\psi$, $\tan\delta\psi\rightarrow\sin\delta\psi$, and $\delta\psi\rightarrow\delta\psi_{\rm 3D}$) are summarized in Fig.~\ref{fig:synComp1}. Here, we plot in log scale the ratio between the DCF-derived magnetic field strength (from Eq.~(\ref{eq:DCF2D})) and the actual mean field strength, $B_{\rm DCF}/\overline{B}$, vs. the mean dispersion of polarization angle $\langle\delta\psi\rangle$, for each model and method. Note that for most of the cases, $\overline{B} = \overline{B_{\rm POS}}$ (and $B_{\rm DCF} = B_{\rm DCF, POS}$), but when considering the projected polarization angles $\delta\psi_{\rm 3D}$, $\overline{B} = \overline{B_{\rm 3D}}$ is used.
Also, two horizontal lines are drawn at $\pm\log(\sqrt{2})$ so that if a  model has all derived values within these two lines, we can claim this model is accurate within a factor of 2. 

We see clear improvement of the accuracy of $B_{\rm DCF}$ when switching from $\langle\delta\psi\rangle$ (top row) to $\delta\psi$ (bottom row), especially for the `vfit' method (left column). With $\delta\psi$, the `vfit' method was able to produce $B_{\rm DCF}$ within a factor of 2 of the actual value for almost all synthetic observations except those with high inclination angles ($\gamma\approx 60^\circ$; red symbols). 
On the other hand, the simple method `eq' (middle column) seems to be less impacted by the choice of $\langle\delta\psi\rangle$ or $\delta\psi$, and the DCF-derived magnetic field strengths using this method are also pretty consistent with the actual values especially for moderate inclination angles ($\gamma\approx 30^\circ$).

To quantitatively compare the accuracy of our proposed methods and revisions to the DCF equation, we define the accuracy measurement $\epsilon$:
\begin{equation}
    \epsilon \equiv \left\langle\left|\log\left(\frac{B_{\rm DCF}}{\overline{B}}\right)\right|\right\rangle\ \ \ {\rm or} \ \ \ \left\langle\left|\log\left(\frac{B_{\rm DCF,POS}}{\overline{B_{\rm POS}}}\right)\right|\right\rangle,
\end{equation}
which is always positive and between 0 and 1. Smaller $\epsilon$ means the estimate of $B_{\rm DCF}$ is more accurate. The mean value of $\epsilon$ for each model is presented in Table~\ref{tab:synobs}. Switching from $\langle\delta\psi\rangle$ to $\delta\psi$ brings $\epsilon$ for `vfit' method down to 0.16 from the original 0.27, and if we added the projection correction and the $\tan\delta\psi\rightarrow\sin\delta\psi$ revision, $\epsilon\approx 0.12$ for `vfit' method, which is the best performing model in this test. 
As mentioned earlier, method `eq' is not affected much by the choices of $\langle\delta\psi\rangle$ or $\delta\psi$. Instead, the accuracy of method `eq' is pretty consistent with $\epsilon\lesssim 0.2$ when using either $\tan\langle\delta\psi\rangle$, $\tan\delta\psi$, or $\sin\delta\psi_{\rm 3D}$ in the DCF equation.

The unexpected result is the bad performance of the `ana' method, which theoretically should give the most accurate estimate of the mean gas density. Instead, the derived $B_{\rm DCF}/\overline{B}$ ratios are very scattered in Fig.~\ref{fig:synComp1}, and none of our proposed revisions helped reduce such error. Combining with the facts that the density estimates from different methods often vary a lot and the fitting method from linewidth-size correlation gives the best estimates, we believe our results suggest that the line-of-sight velocity dispersion used in the DCF equation cannot properly reflect the actual turbulent component of the velocity in the system, and only the mean density value derived also from velocities (the `vfit' method) could offset this uncertainty. Thus, even though the derived mean density is not accurate, the resulting $B_{\rm DCF}$ is closer to the real value than that using the more accurate density from the analytic approach if the line-of-sight velocity dispersion is used, as is commonly the case. 
We note that though the reasoning differs, this is similar to the argument discussed in \cite{Cho_Yoo_2016} that in addition to the commonly-used line-of-sight velocity dispersion, the centroid velocity should also be taken into consideration in the DCF analysis.

Moreover, even though the line-of-sight velocity dispersion is density-weighted, the linewidth could still hugely depend on the low density gas that has larger velocity difference from the density-weighted mean central velocity. Since the mean density is mainly determined by the dense gas, it is less correlated with the velocity dispersion. We thus note that the cloud depth derived from `vfit' method (Eq.~(\ref{eq:vfitL})) should not be treated as the physical cloud depth; rather, it should be viewed as a characteristic length scale corresponding to the measured line-of-sight velocity dispersion. 

The projection effect discussed in Sec.~\ref{sec:2Dproj} is illustrated in Fig.~\ref{fig:synComp1} as solid (POS values) vs.~open (reprojected) symbols. We see that the best scenario for the projection correction to work is the high-inclination low-perturbation case (model L1\_inc60; red squares), which is expected because the mean inclination angle of the magnetic field is more representative in less turbulent environment, and the projection effect is more significant when the field inclination angle is large. We also note that the projection correction seems to work the best with the `eq' density-deriving methods. This could be due to that the `eq' method considers the cloud as an infinite slab with plane-parallel magnetic field (so ${\cal E}_B = B^2/(8\pi)$; see Eq.~(\ref{eq:pres_eq})), so the actual inclination of the magnetic field plays a more critical role in estimating the field strength using this method, especially since the line-of-sight cloud depth also depends on the inclination angle (Eq.~(\ref{eq:Leq})).

Also shown in Fig.~\ref{fig:synComp1} is the modification of the DCF method we proposed in Sec.~\ref{sec:statcorr}, that the term $\tan\delta\psi$ should be replaced by $\sin\delta\psi$ (plotted as single-sided errorbars). We found that though $B_{\rm DCF}$ seems to be slightly more accurate with $\sin\delta\psi$, the improvement is small except for cases with large dispersion of polarization angles ($\langle\delta\psi\rangle \gtrsim 35^\circ$) because $\sin\delta\psi \approx \tan\delta\psi$ for small angles.
While, from the $\epsilon$ values in Table~\ref{tab:synobs}, using $\sin\delta\psi$ is statistically more accurate than the traditional DCF method, the biggest uncertainty in the DCF method seems to be coming from the gas density and turbulence level, and thus the modifications on polarization measurement generally have weak impact on improving the accuracy of the DCF method.





\subsection{Testing Density Dependence and Polarization Selection}
\label{sec:DcutTest}

\begin{table*}
 \begin{center}
  \caption{Summary of the synthetic line observation models discussed in Sec.~\ref{sec:DcutTest}.}
  \label{tab:LineModels}
  \begin{tabular}{lcccccc}
    \hline
    \hline
    model & \multicolumn{2}{c}{$\langle\delta\psi\rangle\ (^\circ)$} & \multicolumn{4}{c}{$\sigma_v$ (km/s)} \\
    & all & $p > 0.2 p_{\rm max}$ & all & $10^2 < n/{\rm cm}^{-3} < 10^4$ & $10^3 < n/{\rm cm}^{-3} < 10^5$ & $10^5 < n/{\rm cm}^{-3}$ \\
    \hline
    L10\_inc0 & 24 & 23 & 0.51 & 0.70 & 0.17 & 0.22 \\
    L10\_inc30 & 28 & 26 & 0.49 & 0.66 & 0.20 & 0.24 \\
    L10\_inc60 & 45 & 44 & 0.75 & 0.83 & 0.27 & 0.22 \\
    \hline
  \end{tabular}
  \end{center}
\end{table*}

\begin{figure*}
\begin{center}
\includegraphics[width=\textwidth]{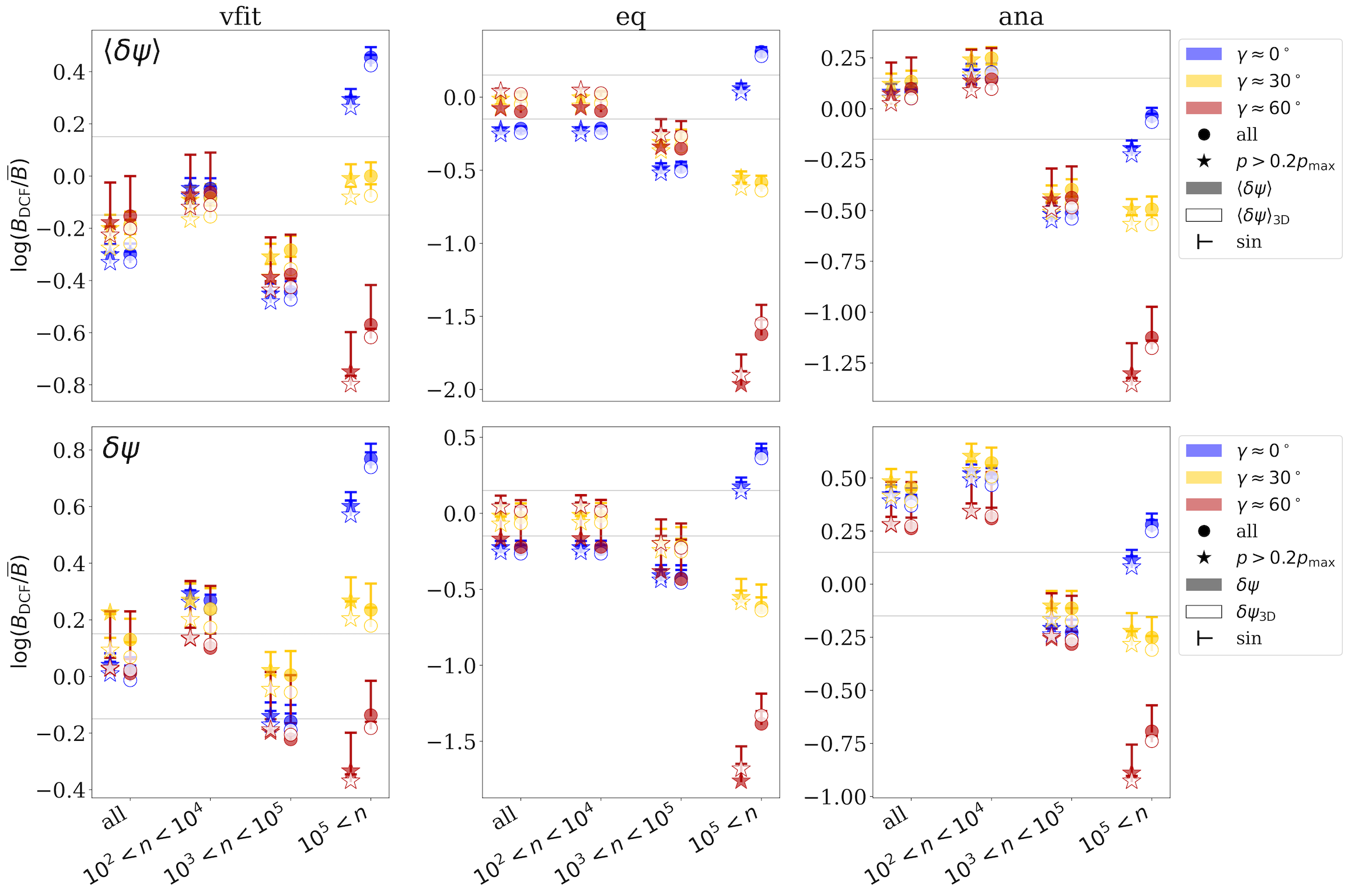}
\vspace{-.2in}
\caption{\small The results from the synthetic line observation tests discussed in Sec.~\ref{sec::SynLine} and~\ref{sec:DcutTest}. Similar to Fig.~\ref{fig:synComp1}, we consider the ratio $B_{\rm DCF}/\overline{B}$ (or $B_{\rm DCF, POS}/\overline{B_{\rm POS}}$) and compare the results from three different density derivations (see Sec.~\ref{sec:rho}) in three different viewing angles of model L10 (see Table~\ref{tab:LineModels}). Also included are the projection-corrected DCF analysis ({\it open symbols}), as well as the polarization selection criterion $p/p_{\rm max} > 0.2$ ({\it stars}) which neglects polarization segments smaller than 20\% of the maximum polarization fraction. The proposed modification to the DCF method using $\sin\langle\delta\psi\rangle$, $\sin\delta\psi$ is also shown as single-sided errorbar. Gray horizontal lines mark $\pm\sqrt{2}$, the boundaries of the factor of 2 accuracy. }
\vspace{-.1in}
\label{fig:NewTest}
\end{center}
\end{figure*}

As discussed in Sec.~\ref{sec::SynLine}, density selection effect could also play a critical role in determining the gas turbulence level and thus the DCF analysis.
Our simplified synthetic line observations were described in Sec.~\ref{sec::SynLine}, and examples of the corresponding column density and velocity dispersion maps were demonstrated in Fig.~\ref{fig:Dcut}.
Here, we consider
three viewing angles of simulation model L10 (synthetic models L10\_inc0, L10\_inc30, and L10\_inc60 in Table~\ref{tab:synobs}) with three different density cutoffs adopted in Fig.~\ref{fig:Dcut} ($10^2 < n/{\rm cm}^{-3}<10^4$, $10^3 < n/{\rm cm}^{-3}<10^5$, and $10^5 < n/{\rm cm}^{-3}$; see Table~\ref{tab:LineModels}) to investigate what density range could provide the most accurate DCF result in star-forming clouds.

In addition, we examine the effect of limiting the analysis to pixels with polarization fraction above a certain threshold ($p > 0.2 p_{\rm max}$ here), as we discussed in Sec.~\ref{sec:synPol}. We would like to point out that this polarization selection effect only has impact on the measured dispersion angle of polarization orientation, while the density selection effect only changes the velocity dispersion. The models considered in this test and the corresponding parameters are listed in Table~\ref{tab:LineModels}.
We note that applying the mask based on polarization fraction does not seem to affect the measured polarization angle dispersion much. On the other hand, the synthetic linewidth hugely depends on the density selection effect, as already suggested in Fig.~\ref{fig:Dcut}. The velocity dispersion tends to be larger in lower-density gas, which also differs from the values measured without any density selection effect (`all' in Table~\ref{tab:LineModels}; column 4). This suggests that any DCF analysis based on synthetic observations without considering the density selection effect may be inaccurate and incompatible to actual observations.

The results of the DCF analysis are summarized in Fig.~\ref{fig:NewTest}. Similar to Fig.~\ref{fig:synComp1}, here we plot the ratio $B_{\rm DCF}/\overline{B}$ ($B_{\rm DCF, POS}/\overline{B_{\rm POS}}$ for cases without projection correction) in log scale, and draw horizontal lines at $\pm \sqrt{2}$ to show the boundaries of accuracy of a factor of 2. 
All values are plotted as functions of the density selection range.
We also include the comparisons between using 1) $\langle\delta\psi\rangle$ and $\delta\psi$ (top vs. bottom row), 2) $\delta\psi$ and $\delta\psi_{\rm 3D}$ (solid vs. open symbols), and 3) $\tan\delta\psi$ and $\sin\delta\psi$ (symbols vs. errorbars) in the DCF equation Eq.~(\ref{eq:DCF2D}).

Interestingly but not surprisingly, the density selection effect does have strong impacts on the accuracy of the DCF result, and the low-to-intermediate density range $10^2 < n/{\rm [cm^{-3}]} < 10^4$ in general delivers the most accurate DCF-derived field strength to the actual value, regardless the field inclination angle with respect to the plane of sky. As discussed above, such discrepancy could be due to the velocity dispersion being sensitive to the density cuts (see e.g.,~Table~\ref{tab:LineModels} and Fig.~\ref{fig:Dcut}). Moreover, the high-density tracers tend to recover only the densest structures, and thus do not have sufficient pixels to have good statistics for the DCF method (see Sec.~\ref{sec:rho} and Fig.~\ref{fig:Dcut}).
The low- and intermediate-density tracers (within the range of $10^2$-$10^5$\,cm$^{-3}$) therefore appear to be the better choices for conducting the DCF analysis in star-forming clouds, which is rational given that this analysis aims to retrieve the cloud-scale magnetic field strength, and $10^2$-$10^5$\,cm$^{-3}$ is indeed the typical range of cloud-scale density.\footnote{We note that it remains an open question that, with enough statistics via high resolution maps, whether the DCF method works in dense, star-forming structures like cores and filaments, in addition to the relatively diffuse and turbulent cloud environment. We shall explore this topic in a following work (J.~Park et al., {\it in prep}). }

We further note that limiting the analysis based on polarization fraction ($p>20\% p_{\rm max}$ adopted here) does not seem to make huge differences (stars vs. circles in Fig.~\ref{fig:NewTest}).
This is consistent to the fact that the mean angle dispersion $\langle\delta\psi\rangle$ values remain similar with or without the polarization fraction limitation, as listed in Table~\ref{tab:LineModels}. 
This may seem surprising, since Fig.~\ref{fig:Bvsp} already showed that the observed polarization orientation and the actual plane-of-sky magnetic field structure are only tightly correlated when the polarization fraction is large enough.
In fact, we repeated the same analysis with the more strict selection criterion $p>50\% p_{\rm max}$, and found that the accuracy of the DCF method decreased instead. This again suggests that the accuracy of the DCF method is not determined by the polarization measurement, but relies on good statistics as well as good hydrodynamics estimates.

Similar to the discussion in the previous section, the projection correction seems to work better with the `eq' method of density derivation, but is only significant when the inclination angle is large (model L10\_inc60; red symbols). Another modification we proposed based on the 3D analysis, $\tan\delta\psi\rightarrow\sin\delta\psi$, does not seem to show consistent improvement to the DCF-derived field strength, which agrees with the results in the previous section.   
Combining with the result that the polarization selection effect does not play a huge role in determining the accuracy of the DCF analysis, we conclude that the DCF analysis relies more on the shape and the peak location of the probability distribution instead of the precise measurement at each location. Conceptually, this agrees with our argument based on our 3D analysis that it is not necessary to have the exact solution of $\hat{\mathbf{B}_0}$ everywhere as long as the distribution of $\delta E_K/\delta E_B$ peaks around 1 (see Sec.~\ref{sec:B0approx}).

\section{Cancellation Effect from Integration}
\label{sec:2Ddisp}

By applying the DCF analysis to their numerical simulations and synthetic polarization measurements, \cite{OSG2001} suggested that a factor of $\xi\approx 0.5$ should be included in the DCF method as a calibration factor:
\begin{equation}
    \langle B\rangle_{\rm DCF, corr} = \xi \langle B\rangle_{\rm DCF} = \xi \sqrt{4\pi\langle\rho\rangle}\frac{\langle\sigma_v\rangle}{\langle\delta\psi\rangle}.
    \label{eq:DCFxi}
\end{equation}
Note that the above DCF equation used in \cite{OSG2001} differs from the commonly-adopted Eq.~(\ref{eq:DCF2D}), because \cite{OSG2001} considered $\langle B_\perp/\overline{B}\rangle \sim \langle\delta\psi\rangle$ instead of $\tan\langle\delta\psi\rangle$ (see \citealt{PSLi_2021} for derivation and justification).
This means that the so-called DCF coefficient of $\xi = 0.5$, based on the results of \cite{OSG2001}, was derived from assuming $B_\perp/B \sim \langle\delta\psi\rangle$, and therefore may not be appropriate when $B_\perp/B \sim \tan\langle\delta\psi\rangle$ is used.
Nevertheless, 
this so-called DCF coefficient $\xi$ has been considered to be a necessary correction factor to acount for the projection/integration effects from inhomogeneities, anisotropies, resolution, and/or variation along the line of sight \cite[see e.g.,][]{Zweibel1990,MG91,Heitsch01}.

However, the $\sin\delta\psi$ vs. $\delta B_{\rm POS}/B_{\rm POS}$ plot in Fig.~\ref{fig:Bvsp} shows that the polarization orientation in our synthetic observations generally follow the projected magnetic field direction pretty well. This suggests that the cancellation effect along the line of sight is not significant in our models, and may explain why we did not need to include the so-called DCF coefficient $\xi \approx 0.5$ as proposed in \cite{OSG2001}. 
More importantly, our results indicate that the DCF coefficient is likely not a constant globally applicable in all environments, and proper calibration is critical in order to increase the accuracy of the DCF method \citep[see recent works e.g.,][]{PSLi_2021,JHLiu_DCF_2021}.
In this section we examine how cancellation along the line of sight could affect the measured polarization angle dispersion, and provide a way to correct for such effect numerically as an extension to the DCF coefficient.


\begin{figure}
\begin{center}
\includegraphics[width=0.5\textwidth]{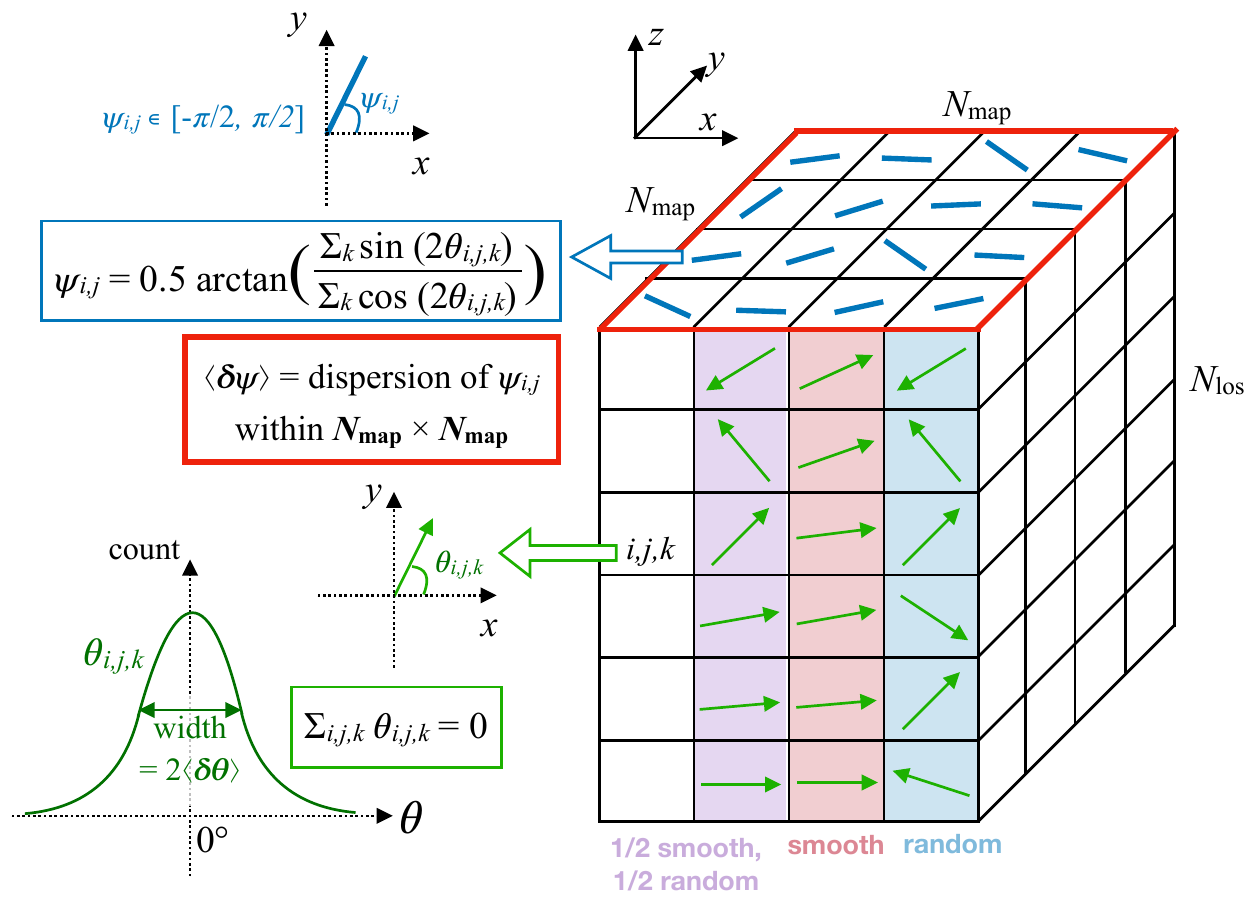}
\vspace{-.1in}
\caption{\small A sketch illustrating the design of the test discussed in Section~\ref{sec:2Ddisp}. We generate a set of $(N_{\rm map}\times N_{\rm map}\times N_{\rm los})$ values of $\theta$, and arrange them into a 3D cube with dimensions $N_{\rm map}\times N_{\rm map}\times N_{\rm los}$ following three different assumptions: smooth, random, and half-smooth, half-random. The synthetic observed angle $\psi$ can therefore be derived from the projection of $\theta$ along each line of sight (a total of $N_{\rm map}\times N_{\rm map}$ values). Comparing the distribution of $\theta$ with that of $\psi$ among different arrangement of $\theta$ can provide insight on how the projection effect could affect the DCF analysis. }
\label{fig:2Ddisp_sketch}
\end{center}
\end{figure}

\begin{figure*}
\begin{center}
\includegraphics[width=\textwidth]{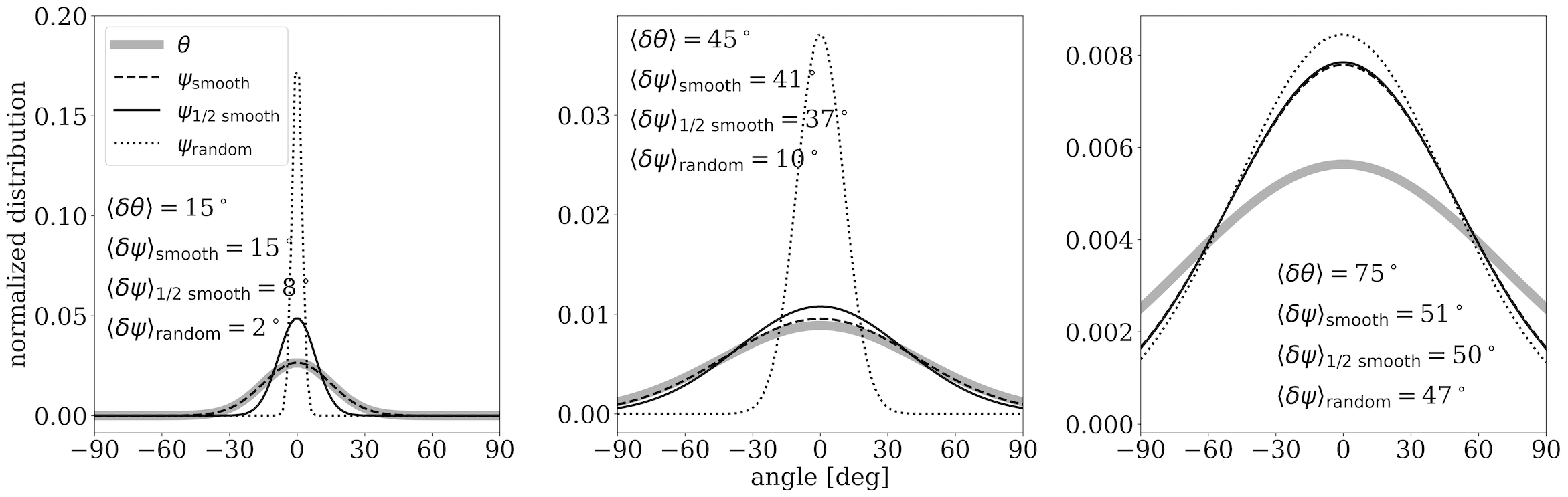}
\vspace{-.2in}
\caption{\small Comparing different arrangement methods of $\theta$ (random, half random, and smooth), using $N_{\rm los}/N_{\rm map} = 0.5$ as an example. Plotted are the normalized distributions of projected polarization angles $\psi$ from the same set of random vectors $\theta$ generated with the corresponding angle dispersion $\langle\delta\theta\rangle$, from the numerical test illustrated in Figure~\ref{fig:2Ddisp_sketch}. This shows that the random arrangement of vectors is the least consistent to the actual distribution of angles, and the smooth arrangement and the half-random, half-smooth case could provide better approximations of the angle dispersion unless the dispersion is very large. }
\label{fig:2DdispPDF}
\end{center}
\end{figure*}


\begin{figure*}
\begin{center}
\includegraphics[width=\textwidth]{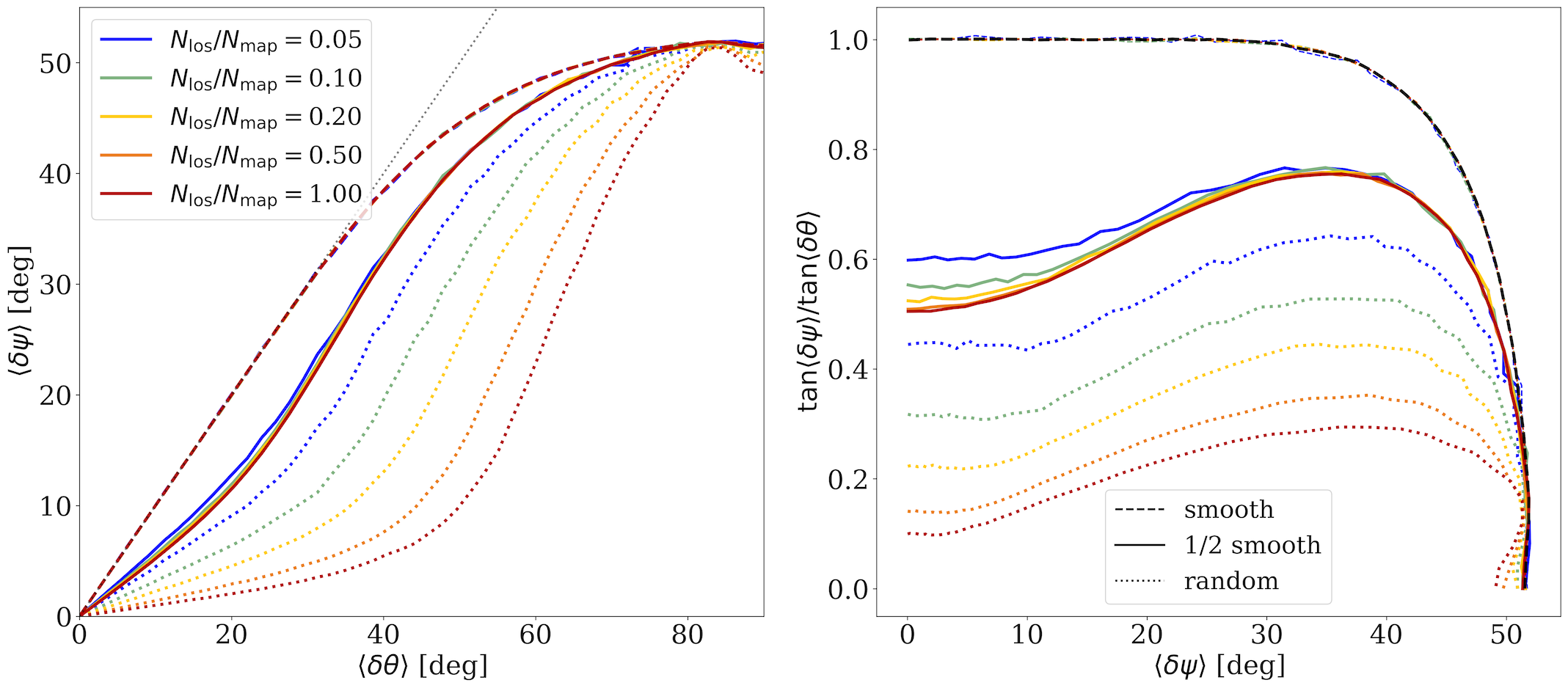}
\vspace{-.2in}
\caption{\small Results from the numerical test illustrated in Figure~\ref{fig:2Ddisp_sketch}, showing the correlation between the dispersion of the projected angle $\langle\delta\psi\rangle$ and the original angle dispersion in 3D $\langle\delta\theta\rangle$, with various ratios between the physical scale on the plane of sky and along the line of sight ($N_{\rm los}/N_{\rm map}$; different line colors) as well as the three different arrangement methods of $\theta$ (different line styles; dashed, solid, and dotted lines represent `smooth', `half-smooth, half-random', and `random' cases, respectively). The correlation between $\langle\delta\theta\rangle$ and $\langle\delta\psi\rangle$ shown on the left panel can be used to derive the correction factor for the DCF analysis, $\tan\langle\delta\psi\rangle/\tan\langle\delta\theta\rangle$ (right panel). }
\vspace{-.1in}
\label{fig:2Ddisp_tan}
\end{center}
\end{figure*}

\begin{figure*}
\begin{center}
\includegraphics[width=\textwidth]{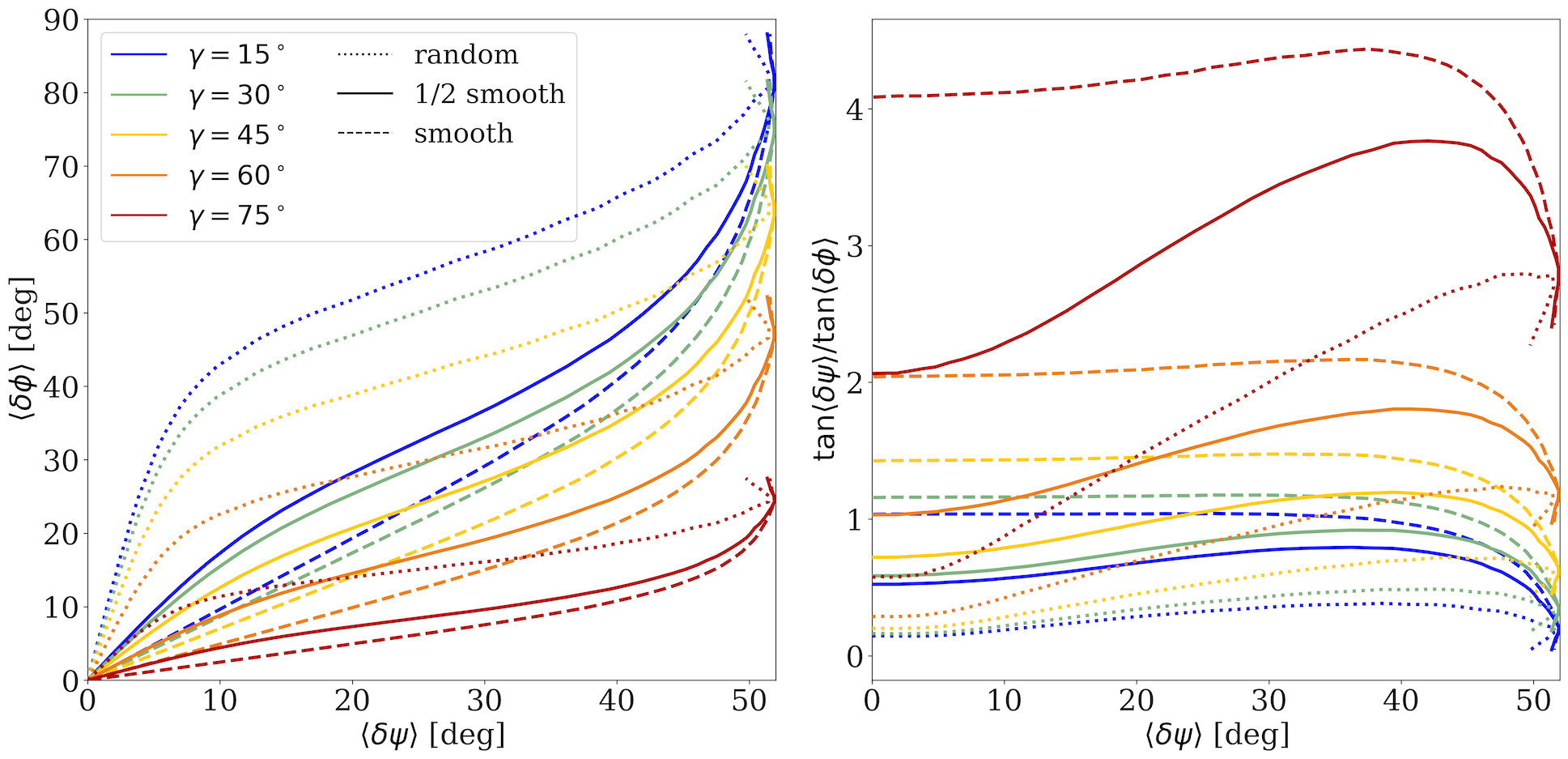}
\vspace{-.2in}
\caption{\small Similar to Fig.~\ref{fig:2Ddisp_tan}, but now also includes the projection effect discussed in Sec.~\ref{sec:2Dproj} (also see Eq.~(\ref{eq:proj3D})). The correlation between the mean angle dispersion measured in 3D, $\langle\delta\phi\rangle$, and the mean dispersion of the synthetic polarization, $\langle\delta\psi\rangle$, depends on both the inclination angle $\gamma$ ({\it different line colors}) and the level of turbulence of the system, which is approximated by different angle assignment method in our numerical test: random ({\it dotted lines}), half random ({\it solid lines}), and smooth ({\it dashed lines}) as described in Sec.~\ref{sec:2Ddisp} and Fig.~\ref{fig:2Ddisp_sketch}.  }
\vspace{-.1in}
\label{fig:obsto3D}
\end{center}
\end{figure*}

Because of the integration along the line of sight, the observed angle dispersion of the projected 2D map could differ from the actual angle dispersion in the 3D space. 
To quantitatively investigate the difference between these two measurements, we designed a numerical test, illustrated in Figure~\ref{fig:2Ddisp_sketch}. 
We first generate a set of 2D angles $\theta$ to fill in a 3D cube with size $N_{\rm map}\times N_{\rm map}\times N_{\rm los}$.  These angles represent the direction of unit vectors on the $x-y$ plane (plane of sky) in each cell.
Note that we ignored the density-dependent weighting (thus unit vectors) to focus on the geometric effect during the integration. 
We also neglected the line-of-sight ($z$-) component of the vector for simplicity (i.e.,~$\theta$ is the 2D projection of the actual 3D vector in each cell); the projection effect for angles from 3D to 2D was discussed in Sec.~\ref{sec:2Dproj}.

The distribution of the generated angle $\theta$ follows a normal distribution centered at $0^\circ$ and with FWHM equal to two times of the chosen dispersion of $\theta$ in the cube, $\langle\delta\theta\rangle$. 
After assigning $\theta$ to each cell to have a 3D array $\theta_{i,j,k}$,
we can now calculate the polarization angle $\psi_{i,j}$ for each column (line of sight) following the general equations of synthetic polarization:
\begin{align}
    u_{i,j} &= \sum_k \sin (2\theta_{i,j,k}),\ \ \ 
    q_{i,j} = \sum_k \cos (2\theta_{i,j,k}),\notag \\
    \psi_{i,j} &= \frac{1}{2}\arctan2(u_{i,j}, q_{i,j}).
\end{align}
The dispersion of polarization angle $\langle\delta\psi\rangle$ can then be derived from the distribution of $\psi_{i,j}$.
The relation between $\psi_{i,j}$ and $\theta_{i,j,k}$ is illustrated in Figure~\ref{fig:2DdispPDF}, which plots the distributions of $\psi$ over the $N_{\rm map}\times N_{\rm map}$ plane under three different cases of $\theta$ assignment. For the case `random', $\theta_{i,j,k}$ is randomly assigned in the box, and thus has the most significant cancellation when integrated along the line of sight. The case `smooth' represent the case when the $\theta$ dataset is sorted before assigning to cells; i.e., the difference in $\theta_{i,j,k}$ along each sightline is minimum. For the `1/2 smooth' case, we randomly assigned half of the dataset but sorted the other half before assigning them to individual cells (also see Fig.~\ref{fig:2Ddisp_sketch} for illustration). Not surprisingly, Fig.~\ref{fig:2DdispPDF} suggests that the `smooth' case best recovers the actual dispersion values unless the real field is really disturbed (large $\langle\delta\theta\rangle)$, while the case of `random' does not seem to be a good approximation even when the angle dispersion is really small.

The correlation between the projected dispersion, $\langle\delta\psi\rangle$, and the dispersion of the whole cube, $\langle\delta\theta\rangle$, is plotted in the left panel of Figure~\ref{fig:2Ddisp_tan}, again for the three cases of $\theta$ assignment. 
In addition, we tested different numbers of pixels along the line of sight as fractions of the size of the 2D map, and as illustrated in Figure~\ref{fig:2Ddisp_tan}, the ratio $N_{\rm los}/N_{\rm map}$ does not have huge impact except for the case where the angles are completely random, which rarely happens in the real world. 
Also note that Figure~\ref{fig:2Ddisp_tan} confirms numerically that the maximum possible value of the measured angle dispersion is $\approx 52^\circ$, as pointed out in \cite{PlanckXIX}.

Our results demonstrate that the projected dispersion is always smaller than the actual dispersion in the cube, which is consistent with what has been suggested due to the cancellation effect along the line of sight \citep[see e.g.,][]{Ostriker2003}. 
Moreover, we can use this numerical test to estimate the corresponding DCF coefficient $\xi$. Assuming $\langle B\rangle_{\rm POS} = \sqrt{4\pi\langle\rho\rangle} {\langle\sigma_v\rangle}/{\tan\langle\delta\theta\rangle}$
holds,\footnote{We note that though we propose to replace $\tan\langle\delta\psi\rangle$ (or $\tan\delta\psi$) with $\sin\langle\delta\psi\rangle$ (or $\sin\delta\psi$) to increase the accuracy of the DCF method, as we discussed in Sec.~\ref{sec:statcorr}, our results in Sec.~\ref{sec::compModel} and~\ref{sec:DcutTest} show that this modification is not critical. We therefore still use the tangent ratio of the polarization angle dispersion for the DCF coefficient.} we have
\begin{equation}
    \langle B\rangle_{\rm POS} = \xi \langle B\rangle_{\rm DCF} = \sqrt{4\pi\langle \rho\rangle}\frac{\langle \sigma_v\rangle}{\tan\langle\delta\psi\rangle},\ \ \ \xi = \frac{\tan\langle\delta\psi\rangle}{\tan\langle\delta\theta\rangle}.
\end{equation}
This ratio is plotted in
Figure~\ref{fig:2Ddisp_tan} (right panel) as a function of the observed angle dispersion on the plane of sky $\langle\delta\psi\rangle$.
Considering the case of `1/2 smooth' (the most realistic assumption among the three considered here),
this shows that if $\tan\langle\delta\psi\rangle$ is used in the DCF method instead of $\tan\langle\delta\theta\rangle$,
the correction factor is ${\tan\langle\delta\psi\rangle}/{\tan\langle\delta\theta\rangle}\approx 0.5$-$0.6$ for small angles.
This is indeed in good agreement with \cite{OSG2001}, who suggested that $\xi\approx 0.5$ for polarization dispersion $\lesssim 25^\circ$.

We note that our numerical test discussed here is similar to the analytical models examined by \cite{MG91}, who also investigated the distribution of angles after integration along the line of sight. 
Also, the modified DCF method using the structure functions of the polarization angles \citep{Hildebrand_09,Houde09} also aims at resolving the cancellation effect along the line of sight due to integration. \cite{Hildebrand_09} and \cite{Houde09} attribute the dispersion of polarization angles to the existence of multiple turbulent cells along the line of sight, which is conceptually similar to our numerical tests: sightlines with more turbulent cells are more similar to the `random' assignment case (see Fig.~\ref{fig:2Ddisp_sketch}) and would have more severe cancellation effect, thus require a correction coefficient further away from 1 (see Fig.~\ref{fig:2Ddisp_tan}), consistent with 
{the method proposed in \cite{Hildebrand_09} and \cite{Houde09} on estimating the ratio between the turbulent component of the magnetic field to the ordered field.}
Further discussion on comparisons between the various modifications of the DCF method will be presented in a separate publication (J.~Park et al., {\it in prep.}).


For completeness, we combine the cancellation effect from integration discussed in this section with the projection effect discussed in Sec.~\ref{sec:2Dproj} to provide the total geometrical correction for the DCF method, summarized in Figure~\ref{fig:obsto3D}. The plot shows the correlation between the measured dispersion $\langle\delta\psi\rangle$ on the plane-of-sky and the actual angle dispersion of the 3D system $\langle\delta\phi\rangle$ (left panel), and the corresponding DCF coefficient $\xi_{\rm 3D} \equiv \tan\langle\delta\psi\rangle / \tan \langle\delta\phi\rangle$ (right panel), for the three cases of angle arrangement (dotted, solid, and dashed lines represent the case of random, half-random, and smooth arrangement, respectively) and various inclination angles $\gamma$. Note that since the ratio $N_{\rm los}/N_{\rm map}$ (i.e.,~the cloud depth relative to the cloud size) does not have a significant impact on the projection and integration effects (unless the angle arrangement is completely random in space), we only plotted the case $N_{\rm los}/N_{\rm map}=0.5$ here. 

Figure~\ref{fig:obsto3D} thus provides a guideline for better estimates of magnetic field strengths using the DCF method with the mean dispersion of polarization angles $\langle\delta\psi\rangle$. 
Note that, unlike the results in Fig.~\ref{fig:2Ddisp_tan} with plane-of-sky angles only, utilizing Fig.~\ref{fig:obsto3D} requires knowing the inclination angle of the magnetic field with respect to the plane of sky, $\gamma$, which can be estimated from the probability distribution function of the observed polarization fraction over the targeted region, as proposed by \cite{Chen2019}. 
Also, though we plotted the three cases of possible magnetic field morphology (different arrangements of $\theta_{i,j,k}$ in our numerical test) along the line of sight, we note that the `smooth' case (dashed lines) should only be considered when the measured $\langle\delta\psi\rangle$ is very small, presumably $\lesssim 10^\circ$. Similarly, the case of completely random arrangement of angles (dotted lines) is unlikely to happen in the real world, and should only be considered if $\langle\delta\psi\rangle$ is closer to the maximum value $\approx 53^\circ$. The `1/2 smooth' case (solid lines) is the most appropriate assumption that one should consider adopting when estimating the 3D correction factor ($\tan\langle\delta\psi\rangle/\tan\langle\delta\phi\rangle$; right panel of Fig.~\ref{fig:obsto3D}) of the DCF analysis.

Nevertheless, we note that the correction factor, or the DCF coefficient $\xi$, is only applicable when considering the dispersion of polarization angle $\langle\delta\psi\rangle$ in the DCF analysis. This is in fact opposite to what we proposed in Sec.~\ref{sec:statcorr} that one should consider $\tan\delta\psi$ instead of $\tan\langle\delta\psi\rangle$, which we showed in Sec.~\ref{sec::compModel} to be more accurate on estimating the field strength. We therefore conclude that one should always consider using $\delta\psi$ in the DCF analysis, and only refer to the correction factor $\tan\langle\delta\psi\rangle/\tan\langle\delta\theta\rangle$ or $\tan\langle\delta\psi\rangle/\tan\langle\delta\phi\rangle$ when the statistics is not good enough to have a log-normal-shaped distribution of the DCF-derived field.


\section{Summary and Conclusions}
\label{sec:sum}

We re-visited the well-known DCF method on deriving magnetic field strength using observed information. Using 3D MHD simulations of star-forming clouds, we tested the fundamental assumption of the DCF method, that the turbulent gas motion is solely responsible for the distortion of the magnetic field morphology, and thus any deviation of the magnetic field from the unperturbed state can be related to the gas velocity through the Alfv{\'e}n wave equation, $\delta E_K = \delta E_B$. While we found that a uniform, unperturbed field did not exist in simulated clouds nor did the strict relation between gas velocity and magnetic field strength, we were able to find a good substitute of the unperturbed field direction: the vector-averaged magnetic field. 
Using the vector-averaged magnetic field as the reference direction,
the ratio between the `perturbed' components (i.e.,~perpendicular to the reference field) of gas kinetic energy and magnetic energy becomes a roughly normal distribution in the log space with the peak around unity. This statistical equipartition between $\delta E_K$ and $\delta E_B$ is the key for the DCF method to work with data in the real 3D space. 

To extend our analysis to observations, we examined the discrepancies between 3D and 2D systems, and proposed several modifications to the original DCF method. These modified DCF methods were then tested using synthetic observations generated from the aforementioned simulations. 
Based on our results, we conclude the best practice of the DCF analysis is the following:

\begin{itemize}
    \item Instead of calculating the DCF-derived field strength using the dispersion of polarization angle $\langle\delta\psi\rangle$ and the mean density and velocity dispersion as $\langle B\rangle_{\rm DCF} = \langle\sigma_v\rangle\sqrt{4\pi\langle\rho\rangle}/\tan\langle\delta\psi\rangle$, we propose to calculate $B_{\rm DCF, local} = \sigma_v\sqrt{4\pi\rho}/\tan\delta\psi$ everywhere on the plane of sky and use the distribution of $B_{\rm DCF, local}$ in log space to find the field strength $B_{\rm DCF} = \langle B_{\rm DCF, local}\rangle_{\rm log}$, which is the location of the peak, or the most probable value of $B_{\rm DCF, local}$ (see Sec.~\ref{sec:statcorr} and Figs.~\ref{fig:AlftestBavg}$-$\ref{fig:Alftest_coremaps}).
    \item The measurement of gas velocity dispersion seems to be critical in the DCF analysis, especially since it can be used in deriving the depth of the cloud and hence the volume density, another essential source of uncertainty in the DCF analysis (see Sec.~\ref{sec::compModel}). However, the velocity dispersion traced by different molecular tracers could be different (see Table~\ref{tab:LineModels}), and our results suggest that the low- to intermediate-density ($\sim 10^3-10^4$\,cm$^{-3}$) tracers are preferred here (see Fig.~\ref{fig:Dcut}). We note that this is the density range that traces the cloud to core transition \citep[see e.g.,][]{CO15}, hence could provide more accurate estimates on the turbulence level of the star-forming gas (see Sec.~\ref{sec:DcutTest}). 
    \item When considering the gas volume density used in the DCF analysis, we recommend to use either the velocity fitting method (Sec.~\ref{sec:vfit}) or the equilibrium layer method (Sec.~\ref{sec:eq}) to derive the characteristic depth of the cloud (see Fig.~\ref{fig:synComp1} and Table~\ref{tab:synobs}), {assuming the region of interest can be considered to be locally flattened.} For the DCF method to work, the gas volume density must be derived from a tracer that is corresponding to the gas traced by polarization measurement, and thus the low- to intermediate-density ($\sim 10^3-10^4$\,cm$^{-3}$) tracers are again the preferred choices (also see Sec.~\ref{sec::SynLine}).
    \item If possible, we recommend using the method proposed in \cite{Chen2019} to estimate the mean inclination angle of the magnetic field with respect to the plane of sky, $\gamma$. If (and only if) $\gamma$ is large ($\gtrsim 60^\circ$), one should consider including the projection correction described in Sec.~\ref{sec:2Dproj} in the DCF analysis by replacing $\delta\psi_{i,j}$ with $\delta\psi_{{\rm 3D}_{i,j}}$ using Eq.~(\ref{eq:proj3D}), and switching $\sigma_{v_{i,j}}$ to $\sigma_{v_{i,j}}\cos\gamma$ to only include the component perpendicular to the inclined magnetic field. The projection effect is minor in most of the cases, but could become critical when the magnetic field is far away from the plane of sky (see Sec.~\ref{sec::compModel}).
    \item The `traditional' DCF method, $\langle B\rangle_{\rm DCF} = \langle\sigma_v\rangle\sqrt{4\pi\langle\rho\rangle}/\tan\langle\delta\psi\rangle$, is only recommended when good statistics on $\delta\psi$ is not available, i.e., when the distribution of $B_{\rm DCF, local} = \sigma_v\sqrt{4\pi\rho}/\tan\delta\psi$ is far off from a log-normal shape. In this case, one should estimate the correction factor $\xi$ using Fig.~\ref{fig:2Ddisp_tan} to get a better approximation of the field strength as $\langle B\rangle_{\rm DCF, corr} = \xi \langle B\rangle_{\rm DCF}$. The projection effect should be considered (Fig.~\ref{fig:obsto3D}) only when the mean inclination angle of the magnetic field $\gamma$ is large.
\end{itemize}

To conclude, good statistics is critical to the DCF analysis, and the biggest uncertainty of the DCF-derived magnetic field strength is actually from the velocity dispersion measurement (especially if the gas volume density is determined using the velocity information), not polarization observation. This explains why neither of our proposed modifications, $\tan\delta\psi \rightarrow \sin\delta\psi$ and adopting a polarization fraction mask $p > 0.2 p_{\rm max}$, has significant impact on the accuracy of the derived magnetic field strength. Nevertheless, our modified DCF analysis should be able to provide estimates of the magnetic field strength within roughly a factor of 2 in typical star-forming clouds except the extreme cases (highly turbulent, large inclination angle, etc.).

\section*{Acknowledgements}

We thank the referee for a constructive and thorough report, and Christopher McKee for productive discussions.
C-YC and Z-YL acknowledge support from NSF grant AST1815784. 
ZYL is supported in part by NASA 80NSSC18K1095. RRM acknowledges support from SOFIA grant 07-0235 and NASA 80NSSC18K0481. 
This research made use of \texttt{astrodendro}, a \texttt{Python} package to compute dendrograms of Astronomical data (\url{http://www.dendrograms.org/}).
This work was performed under the auspices of the U.S.~Department of Energy (DOE) by Lawrence Livermore National Laboratory under Contract DE-AC52-07NA27344 (C.-Y.C and R.I.K.). LLNL-JRNL-832240-DRAFT

\section*{Data Availability}


The simulations used in this work were previously reported in \cite{Chen_HRO_2016,Chen2019,King_2018,King19}. No new data were generated in support of this research. The simulation and synthetic observation data underlying this article will be shared on reasonable request to the corresponding author.



\bibliographystyle{mnras}
\bibliography{reference} 





\bsp	
\label{lastpage}
\end{document}